\documentclass[11pt,a4paper]{article}
\pdfoutput=1
\usepackage{jcappub}
\usepackage[T1]{fontenc}
\usepackage{epsf}
\usepackage{dcolumn}
\usepackage{epsf}
\usepackage{psfrag}
\usepackage{bm}
\usepackage{cleveref}
\usepackage{amstext}
\usepackage{esint}
\usepackage[utf8]{inputenc}

\renewcommand{\vec}[1]{{\bf #1}}       %%%  vectors in bold
\def\beq{\begin{eqnarray}}    %%%  begequation/eqnarray
\def\eeq{\end{eqnarray}}      %%%  endequation/eqnarray
\newcommand{\Om}{\Omega_m}
\newcommand{\Omo}{\Omega_m^0}

\newcommand{\Oro}{\Omega_{r}^0}

\newcommand{\OLo}{\Omega_{\Lambda}^0}

\newcommand{\ODo}{\Omega_{D}^0}

\newcommand{\rc}{\rho_c}
\newcommand{\rco}{\rho_{c}^0}
\newcommand{\rmo}{\rho_{m}^0}

\newcommand{\rmr}{\rho_m}

\newcommand{\rD}{\rho_D}

\newcommand{\rDo}{\rho_{D}^0}

\newcommand{\rL}{\rho_{\CC}}

\newcommand{\rLo}{\rho_{\CC}^0}

\newcommand{\wD}{\omega_D}

\newcommand{\CC}{\Lambda}

\newcommand{\nueff}{\nu_{\rm eff}}

\newcommand{\rRo}{\rho_r^{(0)}}
\newcommand{\be}{\begin{equation}}
\newcommand{\ee}{\end{equation}}
\newcommand{\prm}{\delta\rho_m}
\newcommand{\prD}{\delta\rho_D}
\newcommand{\poD}{\delta\omega_D}
\newcommand{\wDo}{\omega_D^{(0)}}

\newcommand{\DA}{${\cal D}$A}
\newcommand{\DAU}{${\cal D}$A1}
\newcommand{\DAD}{${\cal D}$A2}
\newcommand{\DAT}{${\cal D}$A3}
\newcommand{\DCU}{${\cal D}$C1}
\newcommand{\DCD}{${\cal D}$C2}
\newcommand{\DC}{${\cal D}$C}
\newcommand{\DHlin}{${\cal D}$H}

\title{Background history and cosmic perturbations for a general system of self-conserved dynamical dark energy and matter}

\author[a]{Adri\`a G\'omez-Valent,}
\author[a,b]{Elahe Karimkhani,}
\author[a]{and Joan Sol\`{a}}

\affiliation[a]{High Energy Physics Group, Dept. ECM, and Institut de
Ci\`encies del Cosmos (ICC), Universitat de Barcelona, Av. Diagonal 647,
 E-08028 Barcelona, Catalonia, Spain}

\affiliation[b]{Department of Physics, Faculty of Science, Bu-Ali Sina
University, Hamedan 65178, Iran.}

\emailAdd{adriagova@ecm.ub.edu}
\emailAdd{ e.karimkhani91@basu.ac.ir}
\emailAdd{sola@ecm.ub.edu}

\arxivnumber{1509.03298}

\abstract{
We determine the Hubble expansion and the general cosmic perturbation equations for a general system consisting of self-conserved matter, $\rmr$, and self-conserved dark energy (DE), $\rho_D$. While at the background level the two components are non-interacting, they do interact at the perturbations level. We show that the coupled system of matter and  DE perturbations can be transformed into a single, third order,  matter perturbation equation, which reduces to the (derivative of the) standard one in the case that the DE is just a cosmological constant. As a nontrivial application we analyze a class of dynamical models whose DE density
$\rho_D(H)$ consists of a constant term, $C_0$, and a series of powers of the Hubble rate. These models were previously analyzed from the point of view of dynamical vacuum models, but here we treat them as self-conserved DE models with a dynamical equation of state. We fit them to the wealth of expansion history and linear structure formation data and compare their fit quality with that of the concordance $\CC$CDM model. Those with $C_0=0$ include the so-called ``entropic-force'' and ``QCD-ghost'' DE models, as well as the pure linear model $\rD\sim H$, all of which appear strongly disfavored. The models with $C_0\neq 0$, in contrast, emerge as promising dynamical DE candidates whose phenomenological performance is highly competitive with the rigid $\CC$-term inherent to the $\CC$CDM.
}

\begin{document}
\maketitle
\flushbottom

\section{Introduction}
\label{sec:intro}

It seems now beyond doubt that our Universe is in a state of accelerated expansion owing to some form of dark energy (DE) pervading all corners of the interstellar space. The evidence stems not only from the first and subsequent measurements of distant supernovae \cite{Perlmutter,Riess} but also from the most recent analysis of the precision cosmological data by the Planck collaboration\,\cite{Planck2015}. The physical cause for such positive acceleration is unknown but the simplest and most traditional possibility is to admit the presence of a tiny cosmological constant (CC) in Einstein's field equations,
$\CC>0$. This canonical framework, the so-called ``concordance'' or $\Lambda$CDM
model, seems to describe with good accuracy the available cosmological data
\cite{Planck2015} but, unfortunately, there is at present no special theoretical motivation for it. The CC term in Einstein's equations is usually associated with the energy density carried by the vacuum through the parameter
$\rL=\CC/(8\pi\,G)$ (in which $G$ is the Newtonian coupling). The problem is that it is very difficult to reconcile its value ($\rL\sim 10^{-47}$ GeV$^4$) with the typical expectations in quantum field theory (QFT) and string theory. Such conflict between theory and observation lies at the root of the famous and tough ``old CC problem'' \cite{Weinberg} and its variant
the Cosmic Coincidence problem (see e.g. the reviews \cite{CopSamTsu,Padmanabhan,PeeblesRatra,JSPReview2013}). Both theoretical conundrums stay at the forefront of fundamental physics. Not surprisingly there is plenty of motivation to search for alternative frameworks beyond the $\CC$CDM model.

Needless to say the new proposed models, whatever it be their nature, must be capable of endowing the $\CC$-term of a more satisfactory theoretical status, and at the same time they should be able to maintain (or improve) the quality of the fits to observational data.
Different scenarios have been proposed in order to alleviate this
situation, to wit: scalar fields, both quintessence and phantom-like, modified
gravity theories, phenomenological decaying vacuum models, holography scenarios, etc (cf. the previous
review articles and references therein). In this work, we take the
point of view that the dark energy density is a dynamical variable
in Quantum Field Theory (QFT) in curved spacetime, as in such context it should be
possible to better tackle the basic CC problems\,\footnote{For
recent reviews of the idea of dynamical vacuum energy, see e.g. \cite{JSPReview2013,SolaGomez2015}. See also previous works such as e.g. \cite{BPS09,GSBP11,BasPolarSola12,BasSola2014a,BasSola2014b,GoSolBas2015,GomezSola2015,BasSola2015a} and other recent and related studies such as e.g. \cite{Japonesos,Brasileiros2015,BFLWard,SolaGomezCruz2015}.}. In our approach we do not account for the dynamical character of the DE through specific scalar field models or the like. We rather assume the possibility that the effective form of the DE energy density in QFT in curved spacetime can be expressed as a generic power series of the Hubble rate and its time derivatives.  This possibility can be theoretically motivated from the renormalization group in curved spacetime, see e.g. \cite{JSPReview2013,GRF2015} and references therein.  For the current Universe, however, the relevant terms can only be of order $H^2$ at most, whereas the important possible role played by the higher powers of $H$ in the early Universe can be used to successfully implement inflation, see e.g.\,\cite{SolaGomez2015,GRF2015,LimBasSola2013,BasLimSola2013,LimBasSola2015,BasMavroSola2015}.

The DE models under consideration in this work are closely related to those previously analyzed in detail in the comprehensive studies of Ref. \cite{GoSolBas2015,GomezSola2015}, although with an important difference: now we consider that they describe a self-conserved  dark energy, $\rD(H)$, with a dynamical equation of state (EoS) evolving itself with the expansion history: $\wD=\wD(H)$. It means that $\rD(H)$ does not exchange energy with matter (which therefore remains also covariantly conserved). These assumptions imply a completely different new class of DE scenarios which requires an independent cosmological analysis. We call them the ``${\cal D}$-class'' of dynamical DE models, to distinguish them from the vacuum class (for which $\wD=-1$ at any time of the cosmic expansion). We undertake the task of examining in detail the ${\cal D}$-class here. Of particular interest is the analysis of the effective EoS of the models in this class whose behavior near our time could explain the persistent phantom-like character of the DE without invoking real phantom fields at all.

In this paper we provide detailed considerations not only on the background cosmology of the new DE models but also on the perturbation equations and their implications on the structure formation. The new perturbation equations are indeed formally different from the equations when the models are treated as dynamical vacuum models in interaction with matter. Of special significance is the formal proof that we provide according to which the coupled system of matter and DE perturbation equations for general models with self-conserved DE and matter components can be described by a single, third order, perturbation equation for the matter component. As a nontrivial application we subsequently solve (numerically) that equation for the  ${\cal D}$-class models and compare the results with the situation when the background DE density is present but the corresponding DE perturbations are neglected. The main, and very practical, outcome of our work is that some of these dynamical DE models can provide a highly competitive fit to the overall cosmological data as compared to the performance of the concordance  $\CC$CDM model -- based on a rigid $\CC$-term. For the details of the specific cosmological observables used to fit our models to the data and the procedures of the statistical analysis we provide a  short summary in Sect. 5.1. and refer the reader to the ample exposition previously put forward by some of us in \,\cite{GoSolBas2015,GomezSola2015}.

The layout of the paper is as follows. We present our dynamical DE models in Sect. 2 and address their background solution and equation of state analysis in Section 3. The matter and dark energy perturbations  are considered in Section 4.
The confrontation with the background history and linear growth rate data is performed in Sect. 5. Finally, in Section 6 we present our conclusions.

%%%%%%%%%%%%%%%%%%%%%%%%%%%%%%%%%%%%%%%%%%%%%%%%%%%%%%%%%%%%%%%%%
%%%%%%%%%%%%%%%%%%%%%%%%%%%%%%%%%%%%%%%%%%%%%%%%%%%%%%%%%%%%%%%%%
%%%%%%%%%%%%%%%%%%%%%%%%%%%%%%%%%%%%%%%%%%%%%%%%%%%%%%%%%%%%%%%%%

\section{Dynamical DE models: the \texorpdfstring{$\mathcal{D}$}--class versus the vacuum class}\label{sec:DEmodels}

Let us consider a (spatially) flat FLRW universe.
Einstein's field equations read $G_{\mu\nu}=8\pi G\tilde{T}_{\mu\nu}$, with $G_{\mu\nu}$ the Einstein tensor and $\tilde{T}_{\mu\mu}=T^m_{\mu\nu}+T^D_{\mu\nu}$ the total energy-momentum tensor involving matter and dark energy densities. We take both the matter and DE parts of the energy-momentum tensor in the form of a perfect fluid characterized by isotropic pressures and proper energy densities $(p_i,\rho_i)$. One can then derive Friedmann's equation and the pressure equation by taking
the $00$ and the $ii$ components of Einstein's equations, respectively:
\be\label{eq:FriedmannEq} 3H^2=8\pi\,G\,(\rho_D+\rho_m+\rho_r)\,, \ee
\be\label{eq:PressureEq} 3H^2+2\dot{H}=-8\pi\,G\,(p_D+p_m+p_r)\,, \ee
\noindent where $H=\dot{a}/a$ is the Hubble function and the overdot denotes a derivative with respect to the
cosmic time. As fluid components we consider cold matter, $p_m=0$, radiation, $p_r=\rho_r/3$, and a general dark energy (DE) fluid with dynamical equation of state (EoS): $p_D=\omega_D\rho_D$  ($\dot{\omega}_D\ne 0$). If the matter and DE densities are separately conserved (in the local covariant form which we will indicate explicitly) we shall speak of self-conserved densities. Assuming also that there is no transfer of energy between non-relativistic matter and radiation (which is certainly the case in the epoch under study), the equation of local covariant conservation of the total energy density following from (\ref{eq:FriedmannEq}) and (\ref{eq:PressureEq}) can be split into three equations -- reflecting the Bianchi identities of the Einstein tensor --  as follows:
\be\label{eq:MatterConsEq}
\dot{\rho}_m+3H\rho_m=0\,, \ee
\be\label{eq:RadConsEq}
\dot{\rho}_r+4H\rho_r=0\,,\ee
and
\be\label{eq:DEConsEq}
\dot{\rho}_D+3H\rho_D(1+\omega_D)=0\,. \ee
Similarly, we can e.g. check that from these local conservation laws and the pressure equation (\ref{eq:PressureEq}) we can reconstruct Friedmann's equation \eqref{eq:FriedmannEq}. It follows that as independent set of equations we can take either the pair (\ref{eq:FriedmannEq}) and (\ref{eq:PressureEq}), or Friedman's equation together with the local conservation laws. In the last case the pressure equation (\ref{eq:PressureEq}) is not independent. This will be our strategy in practice.

Trading now the cosmic time derivative for the derivative with respect to the scale factor, through the simple relation $d/dt=aH\,d/da$, the above conservation laws can be easily solved:
\be\label{eq:MatterConsSol}
\rho_m(a)=\rho_m^{(0)}a^{-3}\,, \ee
\be\label{eq:RadConsSol}
\rho_r(a)=\rho_r^{(0)}a^{-4}\,, \ee
\be\label{eq:DEConsEq2} \rho_D(a)=\rho_D^{(0)}\,\exp{\left\{-3\int_{1}^{a}\frac{da^\prime}{a^\prime}[1+\omega_D(a^\prime)]\right\}}\,, \ee
where the densities $\rho_i^{(0)}$ denote the respective current values. In the last case it is assumed that $\wD(a)$ has been computed. However, our method will be actually the opposite, we will compute the DE density $\rD(a)$ first and then use (\ref{eq:DEConsEq2}) to compute the function $\wD(a)$ -- see Eq.\,(\ref{eq:EoS}) below.

As it is transparent from (\ref{eq:MatterConsSol}) and (\ref{eq:RadConsSol}), the pressureless (nonrelativistic) and the relativistic matter energy densities are described by the standard $\Lambda$CDM laws, but the Universe's evolution depends on the specific dynamical
nature of the dark energy density $\rho_D$. The following basic DE models
will be considered\,:
\begin{eqnarray}\label{eq:ModelsA}
\mathcal{D}A1:\phantom{XX} \rho_D(H)&=&\frac{3}{8\pi
G}\left(C_0+\nu H^2\right)\nonumber \\
\mathcal{D}A2:\phantom{XX} \rho_D(H)&=&\frac{3}{8\pi
G}\left(C_0+\nu H^2+\frac{2}{3}\alpha \dot{H}\right)\\
\mathcal{D}A3:\phantom{XX} \rho_D(H)&=&\frac{3}{8\pi
G}\left(C_0+\frac{2}{3}\alpha \dot{H}\right)\nonumber\\
\mathcal{D}C1: \phantom{XX}\rho_D(H)&=&\frac{3}{8\pi G}(\epsilon H_0H+\nu H^2)\nonumber\\
\mathcal{D}C2: \phantom{XX}\rho_D(H)&=&\frac{3}{8\pi G}(\nu H^2+\frac{2}{3}\alpha \dot{H})\label{eq:ModelsC}\,.
\end{eqnarray}
Notice that the constant, additive, parameter $C_0$ has
dimension $2$ (i.e. mass squared) in natural units. We have
introduced the dimensionful constant $H_0$  (the value of the Hubble
parameter at present) as a part of the linear term in $H$ (for \DCU) as in this way the free parameter $\epsilon$ in front of it can be dimensionless.
Similarly $\nu$ and $\alpha$ are dimensionless parameters since they are the coefficients of
$H^2$ and $\dot{H}$, both of dimension $2$. In the case of $\alpha$ we have extracted an explicit factor of $2/3$ for convenience. These free parameters will be fitted to the observational data. For all models we consider two free parameters  at most.

Models \DAU\ and \DAT\ in the list above are, of course, particular cases of \DAD\ corresponding to $\alpha=0$ and $\nu=0$ respectively, but we have given them different labels for convenience and for further reference in the paper.  Similarly, model \DC2\ is, too, a particular case of \DA2\ (for $C_0=0$), but as we shall see it has some particular features that advice a separate study. Model \DC1\,, in contrast, is \textit{not} a particular case of any of the others. A particular realization of the \DC1-model is the case $\nu=0$ (i.e. the purely linear DE model in $H$), which will be denoted
\be
\label{eq:linH}
\mathcal{D}H: \phantom{XX}\rho_D(H)=\frac{3\epsilon H_0}{8\pi G}H\,.
\ee
This ostensibly simple model of the DE has been proposed in the literature on different accounts. It will be analyzed here in detail, along with the rest of the models, to ascertain its phenomenological viability (cf. Sect. 5).

We remark that some of the above models are very similar to the ones we formerly called A1,A2 and C1,C2 in the comprehensive study of Ref.\,\cite{GoSolBas2015}. Formally the expressions for the DE densities are the same, the ``only'' difference being that in the previous reference they were treated as vacuum models (therefore with constant EoS, $\wD=-1$) in interaction with matter, whereas here the effective EoS is a function of the cosmological redshift, $\wD=\wD(z)$, with the additional feature that matter and DE are both conserved.
The ``$\mathcal{D}$'' in front of their names reminds us of the fact that these DE densities will be treated here in the fashion of dynamical DE models with a nontrivial EoS, the latter being determined by the equation of local covariant conservation of the DE density, namely Eq. (\ref{eq:DEConsEq}). We will refer the cosmological DE models (\ref{eq:ModelsA}-\ref{eq:linH}) solved under these specific conditions as the ``${\cal D}$-models'', or the models in the ``${\cal D}$-class'', whereas we reserve the denomination of ``dynamical vacuum models'' when the same DE expressions are solved under the assumption that the EoS is $\wD=-1$ at all times\,\footnote{In this case, in order to fulfil the Bianchi identity, one has to assume that there is an interaction with matter\, \cite{GoSolBas2015} and/or that there is an additional dynamical component (as e.g. in the $\CC$XCDM model\,\cite{LXCDM}), and/or that the gravitational coupling $G$ is running\,\cite{GSBP11,SolaGomezCruz2015}.}. Let us also mention that the case of the pure linear DE model (\ref{eq:linH}) was analyzed in detail in \cite{GomezSola2015} from the point of view of a dynamical vacuum model, but here we will reassess its situation as a ${\cal D}$-model\,\footnote{In analogy with Ref.\,\cite{GoSolBas2015} we could additionally have introduced the ${\cal D}$B$_i$ models, namely the ${\cal D}$-class analogous of the vacuum counterparts B$_i$ defined there. The former are the model types obtained by replacing the $\dot{H}$ term of (\ref{eq:ModelsA}) with the linear term in $H$ when $C_0\neq0$. We will not address here the solution of the general models containing the linear term in $H$ since it is not necessary at this point (cf. \cite{GoSolBas2015,GomezSola2015} for details in the vacuum case). It will suffice for our purposes to study \DC1\ and the pure linear model \DHlin.}.

The cosmological solution of the above ${\cal D}$-models both at the background and perturbations levels turns out to be very different from their vacuum counterparts, as we shall show in this study.
We refer the reader to Ref.\,\cite{GoSolBas2015} for the details on the vacuum part.

A few additional comments are in order before presenting the calculational details. On inspecting the various forms for the ${\cal D}$-models indicated in (\ref{eq:ModelsA}-\ref{eq:linH}), it may be questioned if all these possibilities are theoretically admissible. For
example, the presence of a linear term $\propto H$ in some particular form of $\rD(H)$ -- models \DC1\ and, of course, \DHlin -- deserves some attention. Such term does not respect
the general covariance of the effective action of QFT in curved
spacetime\,\cite{JSPReview2013}. The reason is that it involves
only one time derivative with respect to the scale factor. In
contrast, the terms $H^2$ and $\dot{H}$ involve two derivatives ($\dot{a}^2$ or $\ddot{a}$)
and hence they can be consistent with covariance. From this point of
view one expects that the terms $\propto H^2$ and $\propto\dot{H}$ are primary structures
in a dynamical $\rD$ model, whereas $\propto H$ is not. Still, we
cannot exclude a priori the presence of the linear term since it
can be of phenomenological interest. For example, it could mimic
bulk viscosity effects\,\cite{Japonesos,Viscosity1,Viscosity2}. In addition, some of these models can be viewed also as holographic or entropic-force models\,\cite{Verlinde}, particularly the kind of models of \DC2-type, which were proposed in Ref.\,\cite{Frampton2010}. Let us also mention the connection of the \DHlin\ and \DC-type models with the attempts to understand the DE from the point of view of QCD\,\cite{Schutzhold02}, and the so-called ``QCD ghost dark energy'' models and related ones\,\cite{QCDghost1,QCDghost2,QCDghost3}.  In \cite{BasPolarSola12,BasSola2014b} -- see also \cite{Japonesos} --  these models were shown to be phenomenologically problematic. In fact, models \DC1\,,\DC2\ and \DHlin\ present some phenomenological difficulties which we will elucidate here in the specific context of the ${\cal D}$-class. In this respect let us emphasize that in order to identify the nature of these difficulties it is not enough to judge from the structure of the DE density, e.g. equations (\ref{eq:ModelsA}-\ref{eq:linH}), as the potential problems may reside also in the assumed behavior of matter (e.g. whether matter is conserved or in interaction with the DE). That is why the list of pros and cons of the troublesome models examined in the previous references have to be carefully reassessed in the light of the new assumptions. We will accomplish this task here. As we will see, some of the old problems persist, while others become cured or softened, but new problems also appear. At the same time we will show that the only trouble-free models are indeed those in the large \DA-subclass, i.e. the models possessing a well-defined $\CC$CDM limit. This was shown to be the case as  dynamical vacuum models\,\cite{GoSolBas2015,GomezSola2015} and it will be shown to be so, too, here as ${\cal D}$-models.

\section{Cosmological background solutions}
\label{sec:background}

The first task for us to undertake in order to analyze the above cosmological scenarios is to determine the background cosmological history. From the above equations it is possible, for all the models (\ref{eq:ModelsA}-\ref{eq:linH}), to obtain a closed analytical form for the Hubble function, the energy densities in terms of the scale factor $a$ or, equivalently, in terms of the redshift $z=(1/a-1)$, i.e. $H(z)$. We use them to derive also the EoS and the deceleration parameters, which are very useful to investigate consistency with observational data. Thanks to the self-conservation of dark energy, the EoS can be extracted from the derivative of the DE density with respect to the cosmic redshift,
\be\label{eq:EoS}
\omega_D(z)=-1+\frac{1+z}{3\rho_D(z)}\frac{d\rho_D(z)}{dz}\,.
\ee
Similarly, the deceleration parameter emerges from the corresponding derivative of the Hubble rate, $H$:
\be\label{eq:qz}
q(z)=-1+\frac{1+z}{2H^2(z)}\frac{dH^2(z)}{dz}=-1+\frac{1+z}{2E^2(z)}\frac{dE^2(z)}{dz}\,.
\ee
In the last expression we have used the normalized Hubble rate with respect to the current value, $E(z)=H(z)/H_0$, a dimensionless quantity that will be useful throughout our analysis.
The transition redshift point $z=z_{tr}$ from cosmic deceleration to acceleration can then be computed by solving the algebraic equation
\be\label{eq:zt}
q(z_{tr})=0\,.
\ee
It is important to compute this transition point for the various models, as in some cases there may be significant deviations from the $\CC$CDM prediction.

In the next section we systematically solve the background cosmologies for the \DA- and \DC-type models. The more complicated details concerning  the solution at the perturbations level is presented right next.

\subsection{\texorpdfstring{\DA}--models: background cosmology and equation of state analysis}\label{sect:BackgroundCosmology}

For the general \DAD-model we have to solve the following differential equation:

\be\label{eq:WHEq}
3(1-\nu)H^2=3H_0^2(\Omega_r^{(0)}a^{-4}+\Omega_m^{(0)}a^{-3})+3C_0+\alpha a\,\frac{dH^2}{da}\,,
\ee
which follows from inserting the corresponding expression (\ref{eq:ModelsA}) of the DE density into Friedmann's equation and trading the cosmic time variable for the scale factor. We use also the fact that for the models under consideration the matter is locally self-conserved. The differential equation (\ref{eq:WHEq}) is very different from the one obtained when the model is treated as a vacuum model in interaction with matter (cf. Ref.\,\cite{GoSolBas2015}) and therefore we expect that the \DA-models should have a cosmic history different from the A-type ones.
The constant $C_0$ in (\ref{eq:WHEq}) is fixed by imposing once more the current value of the DE density to be $\rho_D^{(0)}$. This yields
\be
C_0=H_0^2\left[\Omega_D^{(0)}-\nu+\alpha\left(1+\omega_D^{(0)}\Omega_D^{(0)}+\frac{\Omega_r^{(0)}}{3}\right)\right]\,,
\ee
where we have used the relations $\dot{H}_0=-(q_0+1)\,H_0^2$ and
\begin{equation}
q_0=\frac{\Omo}{2}+\Oro+(1+3\wD^0)\,\frac{\ODo}{2}\,,
\end{equation}
with $\wDo$ the EoS value of the DE today. Here the various  $\Omega^{(0)}_i$ stand as usual for the corresponding present day densities normalized with respect to the current critical density $\rco=3H_0^2/(8\pi G)$.
Upon integration of \eqref{eq:WHEq} we obtain the normalized Hubble rate $E(a)=H(a)/H_0$, whose square in this case reads:

\be\label{eq:E2DA2}
E^2(a)=a^{3\beta}+\frac{C_0}{H_0^2(1-\nu)}(1-a^{3\beta})+\frac{\Omega_m^{(0)}}{1-\nu+\alpha}(a^{-3}-a^{3\beta})+\frac{\Omega_r^{(0)}}{1-\nu+4\alpha/3}(a^{-4}-a^{3\beta})\,,
\ee

where $\beta\equiv(1-\nu)/\alpha$. We note the correct normalization $E(1)=1$. It is important to emphasize that we must have $\alpha\geq0$ for these models (in contrast to the situation with the A2-vacuum type considered in \cite{GoSolBas2015}) since otherwise the term proportional to the derivative $dH^2/da$ on the \textit{r.h.s.} of Eq. (\ref{eq:WHEq}) could become arbitrarily large and negative in the past, which would violate the non-negativity of $H^2$ in the corresponding \textit{l.h.s.} of that expression. As a matter of fact, for the models with $C_0\neq0$ (for which $|\nu,\alpha|\ll 1$) we have $\beta\gg1$  since $\alpha$ cannot be negative. Thus, for the term $a^{3\beta}$ appearing in \eqref{eq:E2DA2} we obtain the null effective behavior $a^{3\beta}=(1+z)^{-(1-\nu)/\alpha}\simeq 0$  for most of the cosmic history, namely unless $z$ is extremely close to zero. This observation does not apply for the models \DC\ and \DHlin\,, Eq.\,(\ref{eq:ModelsA}-\ref{eq:linH}), since $C_0=0$ for them and hence the parameters $\nu,\alpha$ can no longer be simultaneously small in absolute value.

For \DA1-type models we take the lateral limit $\alpha\to 0^+$ (i.e. we approach $0$ from the right) in Eq.\,\eqref{eq:E2DA2} (implying $\beta\to+\infty$), from which we can verify that all of the $a^{3\beta}$ terms vanish since in this limit $a^{3\beta}\to 0$ for $a<1$. For $a=1$ (the current time) these terms also cancel because the overall coefficient of $a^{3\beta}$ is $0$ in that limit. The Hubble function becomes, in this case, pretty much simpler:
\be\label{eq:E2DC1}
E^2(a)=1+\frac{\Omega_m^{(0)}}{1-\nu}(a^{-3}-1)+\frac{\Omega_r^{(0)}}{1-\nu}(a^{-4}-1)\,.
\ee
This result for \DA1\ can, of course,  be obtained also from  Eq.\,(\ref{eq:WHEq}), which for $\alpha=0$ just becomes a simple algebraic equation. This shows the internal consistency of the obtained results. For $\nu=0$ we recover of course the $\CC$CDM result.

For illustrative purposes let us write down explicitly the evolution of the DE density in the last case (i.e. for the \DAU-model). Using the redshift variable we find:
\begin{equation}\label{eq:DEa for DA1}
\rD(z)=\frac{\rDo-\nu\rco}{1-\nu}+\frac{\nu}{1-\nu}\left[\rmo\,(1+z)^3+\rRo\,(1+z)^4\right]
\end{equation}
It is easily checked that it satisfies $\rD(0)=\rDo$, as it should, thanks to the cosmic sum rule involving radiation: $\Omo+\Oro+\ODo=1$. And of course it also boils down identically to $\rDo$ for $\nu=0$.

Let us come back to the more general model \DAD. We neglect radiation at this point, as we want to focus now on features of the current Universe, such as the equation of state of the DE near our time. The normalized Hubble function (\ref{eq:E2DA2}) can then be cast in terms of the redshift as follows:
\begin{equation}
\label{eq:hw}
E^2(z)=\frac{C_{0}}{H_0^2(1-\nu)}+\frac{\Omega_{m}^{(0)}}{1-\nu+\alpha}(1+z)^{3}-\eta\left(1+z\right)^{-3\beta}.
\end{equation}
where
\begin{equation}
\label{eq:eta}
\eta=\frac{C_{0}}{H^2_0(1-\nu)}+\frac{\Omega_{m}^{(0)}}{1-\nu+\alpha}-1.
\end{equation}
The evolution of the DE density for the \DA2-models in the matter-dominated and current epoch can be computed with the help of the previous result, yielding
\begin{equation}\label{eq:rDaDA2}
\rD(z)=\frac{\rco\,C_0}{H_0^2(1-\nu)}+\rco\,\Omo\frac{\nu-\alpha}{1-\nu+\alpha}\,(1+z)^{3}-\rco\,{\eta}\,(1+z)^{-3\beta}\,.
\end{equation}
As can be checked, it satisfies $\rD(0)=\rco\,(1-\Omo)=\rDo$ and it identically reduces to $\rDo$ for $\nu=\alpha=0$, i.e. we recover in this limit the $\CC$CDM result. In the last part we use the fact that $(1+z)^{-3\beta}\to 0$ in the limit  $\alpha\to 0^+$ for any $z>0$. One can verify that the \DA1-type solutions \eqref{eq:E2DC1} and \eqref{eq:DEa for DA1} are particular cases of the last results in the matter-dominated and current epochs, as they should. Similarly, the \DA3 -type solution is the particular case obtained from the above formulas for $\nu\to 0$. The numerical evolution of the DE energy densities for these models (normalized to the current value $\rDo$) is shown in Fig.\,1 for the best fit values of Table 1 (see Sect. 5 for the details of the fitting procedure leading to the results of that table). In Figures 2-3 we provide the corresponding behavior of the dark energy EoS and of the deceleration parameter for these models, which will be commented below in turn.

\begin{figure}[!t]
\begin{center}
\includegraphics[scale=0.42]{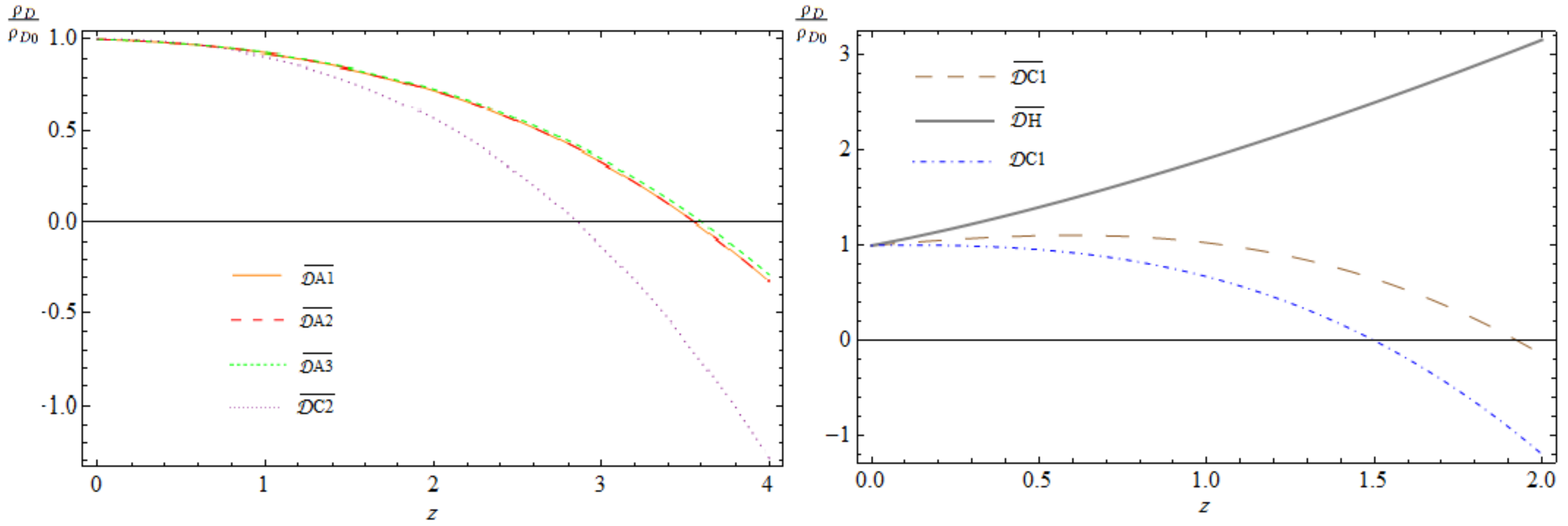}
\caption{\footnotesize{
DE densities normalized to their current value for the various dynamical DE models \DA\ and \DC\ under consideration. The behavior of \DC1\ is different and has been plotted apart, together with the
pure linear model \DHlin. We note that in all plots (unless stated otherwise) we use the best fit values of Table 1 corresponding to the barred quantities, i.e. those obtained when the structure formation data have also been taken into account in the fit.
\label{fig:DE density}}
}
\end{center}
\end{figure}

From \eqref{eq:rDaDA2} and  (\ref{eq:EoS}) we may derive
the EoS for the general subclass of \DA2\ models. It can be expressed in the following compact form:
\begin{equation}
\label{eq:wdw}
\omega_{D}(z)=-\frac{\rco\,C_0}{H_0^2(1-\nu)\rD(z)}+\rco\,\frac{\eta(1+\beta)}{\rD(z)}(1+z)^{-3\beta}\,,
\end{equation}
with $\rD(z)$ given by (\ref{eq:rDaDA2}). For $C_0\neq0$ the parameters $\nu$ and $\alpha$ are small (this is confirmed by the fitting values in Table 1) and therefore we can apply a similar argument to the limiting situation just explained above to prove that the second term on the \textit{r.h.s.} of \eqref{eq:wdw} does not contribute significantly except in the origin. In practice, for any $z>0$ (even for points very close to the origin) the effective EoS  is actually given by the first term on the \textit{r.h.s.} of \eqref{eq:wdw}. This is indeed the result that is connected by continuity with the EoS of the  \DA1-models in the limit $\alpha\to 0^+$, as we shall show in a moment below. Therefore, for all practical situations related to redshift points around our current epoch we establish as effective EoS for the \DA2-models the following expression:
\begin{equation}
\label{eq:EoSDA2}
\omega_{D}(z)=-\frac{1}{1+\frac{H_0^2(1-\nu)}{C_0}\Omo\frac{\nu-\alpha}{1-\nu+\alpha}\,(1+z)^3}\,,
\end{equation}
where by the same token we have disregarded the last term of Eq.\,(\ref{eq:rDaDA2}). There is however one proviso related to the fact that $\rD(z)$ in the denominator of the two terms in (\ref{eq:wdw}) could vanish, and in fact does vanish in our case for some particular redshift value (see below). This causes the presence of a vertical asymptote at a finite $z$ point.  In these cases, one would think of using Eq. (\ref{eq:wdw}) to better describe the behavior around the asymptote.  In actual fact, not even this possibility affects in any significant way the practical use of Eq.\, \eqref{eq:EoSDA2}, as we have checked.
\begin{table}
\begin{center}
\resizebox{1\textwidth}{!}{
\begin{tabular}{| c  |c | c | c | c | c | c |c | c | c  | c |c |}
\multicolumn{1}{c}{Model}  & \multicolumn{1}{c}{$\Omega_m^{(0)}$} & \multicolumn{1}{c}{$\overline{\Omega}_m^{(0)}$}  &  \multicolumn{1}{c}{{\small$\nueff=\nu-\alpha$} }  & \multicolumn{1}{c}{{\small$\bar{\nu}_{\rm eff}$}}    & \multicolumn{1}{c}{$\sigma_{8}$} & \multicolumn{1}{c}{$\overline{\sigma}_{8}$} &
\multicolumn{1}{c}{$\chi^2_{r}/dof$}  &
\multicolumn{1}{c}{$\chi^2/dof$} &
\multicolumn{1}{c}{$\overline{\chi}^2/dof$} &
\multicolumn{1}{c}{AIC} &
\multicolumn{1}{c}{$\overline{\rm AIC}$}
\\\hline {\small $\CC$CDM}  & {\small$0.291^{+0.008}_{-0.007}$} & {\small$0.286\pm 0.007$} & - & -  & {\small$0.815$} & {\small$0.815$} & {\small$569.21/592$} & {\small$584.91/608$} & {\small$584.38/608$} & {\small$586.91$} & {\small$586.38$}
\\\hline
{\small $\mathcal{D}$A1}  & {\small$0.286^{+0.012}_{-0.011}$} & {\small$0.281\pm 0.005$} & {\small$-0.024\pm 0.018$} & {\small $-0.028\pm 0.016$} & {\small$0.773$} & {\small$0.770$} & {\small$565.50/591$} & {\small$573.02/607$} & {\small$573.31/607$} & {\small$577.02$} & {\small$577.31$}
\\\hline
{\small $\mathcal{D}$A2}  & {\small$0.286^\pm 0.011$} & {\small$0.281\pm 0.005$} & {\small$-0.024\pm 0.018$} & {\small $-0.028\pm 0.016$} & {\small$0.772$}  & {\small$0.769$} & {\small$565.57/591$}& {\small$573.03/607$} & {\small$573.40/607$} & {\small$577.03$} & {\small$577.40$}
\\\hline
{\small $\mathcal{D}$A3}  & {\small$0.287^\pm 0.011$} & {\small$0.282\pm 0.005$} & {\small$-0.023^{+0.017}_{-0.018}$} & {\small $-0.027\pm 0.015$} & {\small$0.777$} & {\small$0.773$} & {\small$565.63/591$} & {\small$573.44/607$} & {\small$573.47/607$} & {\small$577.44$} & {\small$577.47$}
\\\hline
{\small $\mathcal{D}$C1}  & {\small$0.286\pm 0.014$} & {\small$0.335\pm 0.007$} & {\small$-0.64\pm 0.13$} & {\small$-0.35\pm 0.05$} & {\small$0.440$}   & {\small$0.735$} & {\small$563.86/584$} & {\small$880.74/600$} &{\small$635.23/600$} & {\small$884.74$} & {\small$639.23$}
\\\hline
{\small \DHlin}  & {\small$0.242\pm 0.008$} & {\small$0.286\pm 0.005$} & - & - & {\small$0.513$} & {\small$0.729$} & {\small$639.85/585$} & {\small$809.61/601$} & {\small$677.11/601$} & {\small$811.61$} & {\small$679.11$}
\\\hline
{\small $\mathcal{D}$C2}  & {\small$0.285\pm 0.013$} & {\small$0.295\pm 0.006$} & {\small$1.03^{+0.09}_{-0.06}$} & {\small$1.02\pm 0.01$} & {\small$0.666$}   & {\small$0.752$} & {\small$563.53/584$} & {\small$594.13/600$} & {\small$572.17/600$} & {\small$598.13$} & {\small$576.17$}
\\\hline
 \end{tabular}}
\end{center}
\caption{\scriptsize The best-fitting values for the various models and their
statistical  significance ($\chi^2$-test and Akaike information criterion AIC\cite{Akaike}, see Sect. \ref{sect:Fitting}).
All quantities with a bar involve a fit to the total input data, i.e. the expansion history (BAO$_A$+BAO$_{d_z}$+SNIa) and CMB shift parameter data, as well as the linear growth data. Those without bar correspond to a fit in which we use all data but exclude the growth data points from the fitting procedure. The value $\chi^2_{r}$ is the reduced $\chi^2$, which does not include the linear growth $\chi^2_{f\sigma_8}$ contribution.  For models \DA1 (resp. \DA3) $\nueff=\nu$ (resp. $-\alpha$); for \DA2\ we have fixed $\alpha=-\nu$ to break degeneracies (see text). For \DC\ and \DHlin\ models we have not used the BAO$_{d_z}$ and CMB data  for the reasons explained in the text. In addition, for the \DC\ models the given value of $\nueff$ must be understood as the value of $\nu$ since $\alpha$ is not defined for \DC1\ and becomes determined for \DC2  (see text). The quoted number of degrees of freedom ($dof$) is equal to the number of data points minus the number of independent fitting parameters.
%Models A1, A2 are referential vacuum models of \,\cite{GoSolBas2015}, which we have re-fitted here under the same conditions to ease comparison with their \DA\ counterparts.
Details of the fitting observables are given in Sect. \ref{sect:Fitting}. \label{tableFit2}}
\end{table}
Being the parameters $\nu$ and $\alpha$  small in absolute value  we can expand the expression (\ref{eq:EoSDA2}) linearly in them. The result can be cast in the following suggestive form for redshift points near our time (typically valid in the more accessible region  $0<z\lesssim2$):
\begin{equation}\label{eq:EoS approx}
\omega_{D}(z)\simeq-1+\frac{H_0^2(1-\nu)}{C_0}\,\Omo\,\nueff\,(1+z)^3\simeq -1+\frac{\Omo}{1-\Omo}\,\nueff\,(1+z)^3\,.
\end{equation}
In this expression, the dimensionless quantity
\begin{equation}\label{eq:nueff}
\nueff=\nu-\alpha
\end{equation}
is the basic fitting parameter. Using its value and that of $\bar{\Omega}_m^{(0)}$ from Table 1 we can evaluate the current EoS parameter of the \DA2\ model, Eq.\,(\ref{eq:EoS approx}):
\begin{equation}\label{eq:EoScurrentDA2}
\omega_{D}^{(0)}\equiv \omega_D(0)\simeq -1.011\,,
\end{equation}
which turns out to be remarkably close to the $\CC$CDM behavior.
Let us also stress at this point that $\nueff$ is indeed the effective fitting parameter for the entire set of \DA-models at the background level in the matter-dominated and current epochs. It is a small parameter since $|\nu,\alpha|\ll 1$ (owing to $C_0\neq 0$). In particular, let us note that because $\omega_{D}^{(0)}$ is very close to $-1$, the coefficient carrying $C_0$ in the Hubble function (\ref{eq:E2DA2}) can be written in first order as $\ODo-\Omo\nueff$ in the limit $\omega_{D}^{(0)}\to -1$. On the other hand the remaining $\nu$ and $\alpha$-dependence in $\beta$ virtually disappears for the reasons discussed above.

In the radiation epoch, however, the dependence on $\nu$ and $\alpha$ is different than (\ref{eq:nueff}), as it is clear from the radiation term in \eqref{eq:E2DA2}. Also, in Sect. 5 we will see that this produces corrections to the transfer function which depend separately on $\nu$ and $\alpha$, and in these cases we must unavoidably fix some relation between these parameters. It has been indicated in the caption of Table 1, and more details are given in Sect. 5.

There is one more, worth noticing, feature to stand out in connection to the numerical value (\ref{eq:EoScurrentDA2}) and the effective EoS function \eqref{eq:EoS approx}. They suggest that the dynamical DE models under study can provide, in principle, a reason for the quintessence and phantom-like character of the DE without necessarily using fundamental scalar fields. Indeed, we find that for $\nu>\alpha$ (i.e. $\nueff>0$) the \DA-models behave effectively as quintessence whereas for $\nu<\alpha$ (i.e. $\nueff<0$) they behave phantom-like near our time. In practice, after fitting the overall cosmological data, we have seen that the latter is the observed situation and also the predicted theoretical result. Recall that the current observational evidence on the dark energy EoS, for example from Planck results, leads to $\omega_{D}^{(0)}=-1.006\pm 0.045$\,\cite{Planck2015}. This is perfectly compatible with the estimate (\ref{eq:EoScurrentDA2}) both quantitatively and qualitatively. This said, we are not suggesting that the current data on the EoS of the dark energy implies that the DE is phantom-like, as the accuracy around the central value is still insufficient. However, for years the central value of the $\wD^{(0)}$ measurement has shown some tilt into the region below $-1$, even during the long period of WMAP observations\,\cite{WMAP1,WMAP2}. Here we are merely saying that if such kind of measurement would be reinforced in the future, the general class of ${\cal D}$-models encodes the ability for providing an explanation of the phantom character of the observed DE without need of invoking real phantom fields at all.  Being $\nueff={\cal O}(10^{-2})$ rather small (and negative) the departure of $-1$ from below is very small, which is precisely the kind of result compatible with observations. To produce the plots of Fig. 2 (left) we have used both the exact expression (\ref{eq:wdw}) for the effective EoS of the \DA2-model and the effective one (\ref{eq:EoSDA2}), and have found no appreciable numerical differences.

\begin{figure}[!t]
\begin{center}
\includegraphics[scale=0.42]{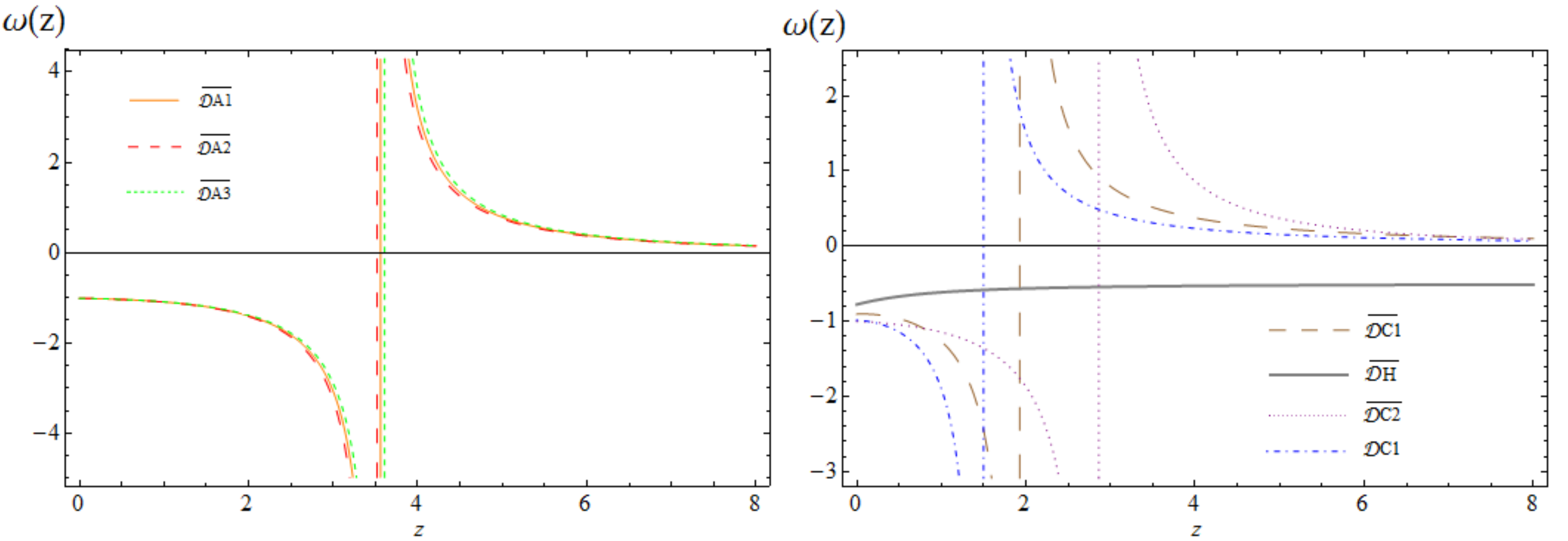}
\caption{\footnotesize{
EoS function $\omega_{D}(z)$ for the DE models \DA\  (left) and for  \DC1\ and \DC2\ (right).  The vertical asymptotes are located at the points where the DE density vanishes (compare with Fig.\,1). The linear \DHlin\ model is seen not to present any asymptote (cf. right plot).
\label{fig:w(z)z1}}
}
\end{center}
\end{figure}

Let us now say some words on the presence of vertical asymptotes in Fig. 2. All models under study display these asymptotes, except \DHlin\ (cf. next section). Such pole-like singularity is observed also in other contexts, for example in non-parametric reconstructions of the EoS function $\wD(z)$\,\cite{Shafieloo2006}, in certain brane-model cosmologies\,\cite{Sahni2003} and in other situations, such as e.g. in mimicking quintessence and phantom DE through a variable $\CC$ \cite{SS0506,BasSola2014a}. In our case the pole-like feature is related to the fact that the denominator of (\ref{eq:EoSDA2}) vanishes for some finite value of $z$, which is to be expected since $\nueff<0$ (cf. Table 1). Physically this means that the DE density $\rD(z)$ vanishes at these values of $z$ (around $3.5$ for models \DA\,, and near $1.5$ or $2$ for \DC1\ depending on the fit options indicated in Table 1), as it is confirmed from the behavior of $\rD(z)$ in Fig\,1. As a consequence of this fact the EoS function (\ref{eq:EoSDA2}) develops a singularity at this point. Notice, however, that the late-time expansion
of the universe in the wide span $0<z<2$ (possibly comprising all relevant supernovae data) is free from these exotic behaviors, if using the most optimized fit values that include the structure formation data. Obviously the latter are not associated with inherent pathologies of the model since the values of the fundamental physical quantities, such as the energy densities, stay finite (e.g. $\rD(z)$ simply vanishes at these ``singular'' points). Interestingly, the very existence of these points might carry valuable information, for if the DE would be described by the ${\cal D}$-models and we could eventually explore the EoS behavior at high redshifts ($z>2$) we should be able to pin down these features, which would manifest through an apparent flip from phantom-like behavior into quintessence-like one when observing points before and after, respectively, the one where the DE density vanishes (cf. Fig. 1). Observing such phenomenon could be a spectacular signature of these models\,\footnote{It should be stressed that the present situation is different from the EoS studies previously entertained in Refs.\,\cite{SS0506,BasSola2014a}, in the sense that in the latter the effective EoS was only a representation (the so-called ``DE picture'') of the original vacuum model (which in turn was called the ``CC-picture''\,\cite{SS0506}), whereas here the EoS under study stands for the ``physical'' one associated with the original  ${\cal D}$-model. Therefore, if these models are to provide a correct description of the DE we should be able to observe the exotic EoS patterns shown in Fig.\,2, much in the same way as in the alternative frameworks proposed in Refs.\, \cite{Shafieloo2006,Sahni2003}.}.

\begin{figure}[!t]
\begin{center}
\includegraphics[scale=0.42]{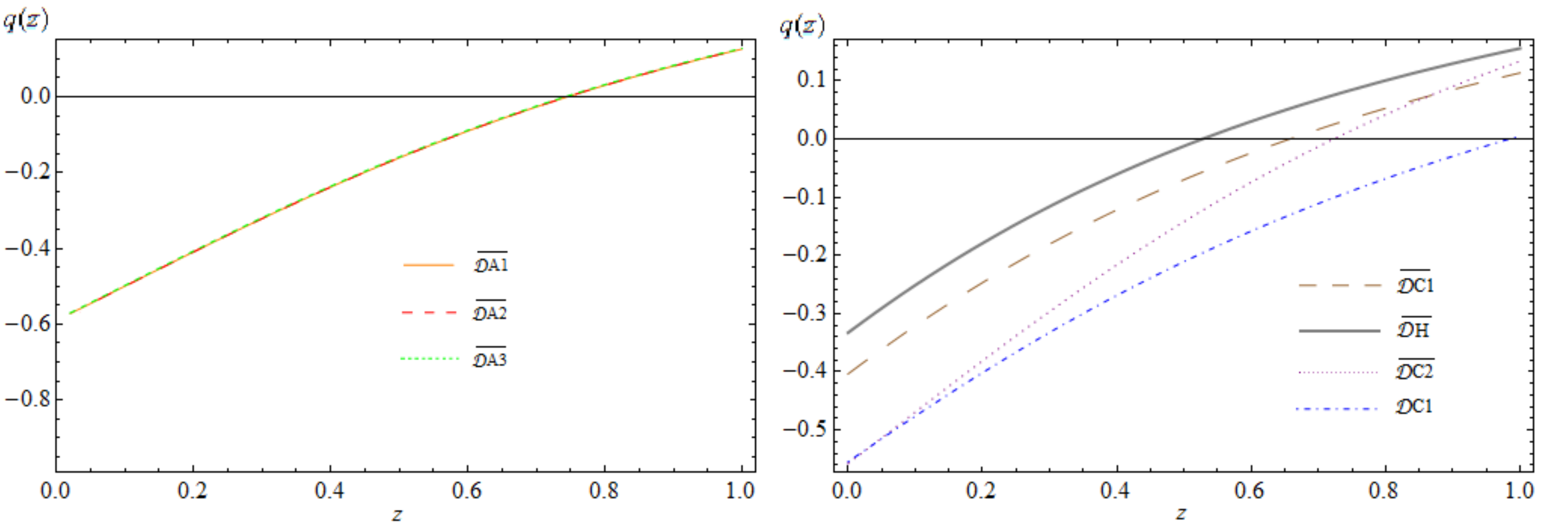}
\caption{\footnotesize{
Deceleration parameter, $q(z)$, and transition point from deceleration to acceleration for the various \DA\ and \DC\ models. Notice that the transition point $q(z_{tr})=0$ from deceleration to acceleration changes significantly for \DC1-models depending on whether we use barred or unbarred fitting parameters in Table 1. In contrast, \DA\ and \DC2-models have a more similar transition redshift $z_{tr}$ which is not far away from that of the $\CC$CDM.
\label{fig:q(z)}}
}
\end{center}
\end{figure}

The EoS for the simpler subclass of \DA1-models can now be obtained from the limit $\alpha\to 0^+$ of Eq.\,(\ref{eq:wdw}). In this limit the second term on the \textit{r.h.s} exactly cancels for any $z>0$ and we are left only with the first term. The result is simply Eq.\,(\ref{eq:EoSDA2}) for $\alpha=0$. In particular, we can see that for $z\to 0$ we obtain a prediction for the current EoS value for this model:
\be\label{eq:omegaDo for DA1}
\omega_D^{(0)}=-\left[\frac{1-\Omega_m^{(0)}-\nu}{(1-\nu)(1-\Omega_m^{(0)})}\right]\simeq -1+\nu\,\frac{\Omo}{1-\Omo}\,,
\ee
where the first expression is exact and the second is valid for $|\nu|\ll1$. The latter is seen to be consistent with the limit $z\to 0$ of Eq.\,(\ref{eq:EoS approx}). Since in this case the fitting values of $\nu$ and $\nueff$ in Table 1 are the same, we retrieve the numerical result (\ref{eq:EoScurrentDA2}) also for \DA1\,. Interestingly, we can explicitly verify that the result (\ref{eq:omegaDo for DA1}), which we have first obtained by taking  the limit $\alpha\to 0^+$ in Eq.\,(\ref{eq:wdw}), can also be worked out from (\ref{eq:EoS}) using the specific Eq.\, (\ref{eq:DEa for DA1}) of the \DA1-model.

The deceleration parameter for \DA-type models reads:
\begin{equation}
\label{eq:qw}
q(z)=-1+\frac{1}{2E^{2}(z)}\left(\frac{3\Omega_{m}^{(0)}}{1-\nu+\alpha}\left(1+z\right)^{3}+3\beta \eta\left(1+z\right)^{-3\beta}\right)\,.
\end{equation}
Solving Eq.\, (\ref{eq:zt}) in this case we may find the transition redshift for a general \DA2-type model. Once more we neglect the last term of Eq.\eqref{eq:qw} since it gives a negligible correction, and we arrive at

\be\label{eq:ztrDA2}
z_{tr}=\left[\frac{2(1+\alpha-\nu)[1-\Omega_m^{(0)}-\nu+\alpha(1+\omega_D^{(0)}\Omega_D^{(0)})]}{\Omega_m^{(0)}(1-\nu)}\right]^{1/3}-1\,.
\ee
The corresponding result for \DA1\ is obtained by setting $\alpha=0$ in the above expression. Numerically, the deceleration and  EoS parameters at the current time for $\mathcal{D}$A1  read respectively as follows: $q^{(0)}=-0.590$ and $z_{tr}=0.745$. For $\mathcal{D}$A3 models, the results are $q^{(0)}=-0.589$ and $z_{tr}=0.742$. For $\nu=\alpha=0$ the formula (\ref{eq:ztrDA2}) naturally returns the $\CC$CDM result:
\begin{equation}\label{eq:ztrLCDM}
z_{tr}=\left(\frac{2\ODo}{\Omo}\right)^{1/3}-1\,,
\end{equation}
whose numerical value for the fitting parameters in Table 1 is $z_{tr}\simeq 0.709$.  The plot of $q(z)$ for the \DA\ models is depicted in Fig. 3 (left), where we can also read the transition point $z_{tr}$.

%and $q^{(0)}=-0.368$ and $z_{tr}=0.766$ for W3 models. But in the W3 models there is another transition point which could be seen in Fig.(\ref{fig:q(z)W}) at $z=-0.01$.

\subsection{Models \texorpdfstring{$\mathcal{D}$}{DC1}C1, \texorpdfstring{$\mathcal{D}$}{DH}H and \texorpdfstring{$\mathcal{D}$}{DC2}C2}
\label{sect:DCModels}

Let us start with \DCU. The normalized Hubble rate $E(z)= H(z)/H_0$ can easily be found starting from Friedman's equation upon inserting in it the corresponding DE density from Eq.\,\eqref{eq:ModelsC}:
\be\label{eq:HubbleDC1}
E(z)=\frac{\epsilon+\Sigma(z)}{2(1-\nu)}\,,\ee
with
\be
\Sigma(z)=\sqrt{\epsilon^2+4(1-\nu)[\Omega_r^{(0)}\,(1+z)^{4}+\Omega_m^{(0)}(1+z)^{3}]}\,
\ee
The parameters in the above relation are constrained to satisfy $1-\nu=\epsilon+\Omega_m^{(0)}+\Omega_r^{(0)}$ so as to insure that $E(0)=1$. Equivalently, $\epsilon=\ODo-\nu$. Thus we can take e.g. $\nu$ as the single free model-parameter, given the values of the ordinary parameters $\Omega_m^{(0)}$ and $\Omega_r^{(0)}$. For this model indeed, $\nueff$ in Table 1 is meant to be $\nu$. Furthermore, the same constraint shows that $\epsilon$ and $\nu$ cannot be both small parameters (in absolute value) in the case of the \DCU\ models, i.e. they do \emph{not} satisfy $|\epsilon,\nu|\ll 1$, in contrast to ${\cal D}$A models, the reason being that the \DCU\ models do not have a well-defined $\CC$CDM limit for any value of $\epsilon$ and $\nu$. As a result one of the two parameter can be of order $\mathcal{O}(1)$. For example, the fit in Table 1 for the barred quantities indicates that $\bar{\nu}\simeq -0.35$ and $\bar{\Omega}_m^{(0)}\simeq 0.335$, and therefore since the radiation component is negligible at present we obtain $\epsilon\simeq 1.02$. For the case when the structure formation data are not used in the fit (unbarred parameters) we have ${\nu}\simeq -0.64$, which is much larger in absolute value, and then $\epsilon$ is also larger: $\epsilon\simeq 1.35$.

The corresponding expression for the DE density of this model reads
\be\label{eq:rDforDC1}
\rD(z)=\rco\,\left[\epsilon\,E(z)+\nu\,E^2(z)\right]\,,
\ee
where $E(z)$ is given by (\ref{eq:HubbleDC1}). At $z=0$ the above function correctly renders the DE density now: $\rD(0)=\rco\,(\epsilon+\nu)=\rco\,(1-\Omo-\Oro)=\rco\,\ODo=\rDo$ upon using the mentioned constraint between the parameters of the model.
For considerations in the matter-dominated and current epochs we can ignore of course the radiation component. A plot of the function (\ref{eq:rDforDC1}), normalized to its current value $\rDo$, is shown in Fig. 1 (right).  In the same figure we plot the case of \DC1\ when $\nu=0$, i.e. the linear model \DHlin, Eq.\,(\ref{eq:linH}). This model has no free parameter, apart from $\Omo$, since the above mentioned constraint enforces the relation  $\epsilon=1-\Omo=\ODo$ in the matter-dominated epoch. Therefore,
\begin{equation}\label{eq:EzDH}
E(z)=\frac12\,\left(\ODo+\left.\Sigma(z)\right|_{\nu=0}\right)=\frac12\,\left(\Omega_D^{(0)}+\sqrt{\Omega_D^{(0)2}+4\Omo\,(1+z)^3}\right)\,.
\end{equation}
The corresponding DE density is:
\begin{equation}\label{eq:rDzDH}
\rD(z)=\rco\,\ODo\,E(z)=\frac12\,\rDo\left(\Omega_D^{(0)}+\sqrt{\Omega_D^{(0)2}+4\Omo\,(1+z)^3}\right)\,.
\end{equation}
Notice the correct normalization $\rD(0)=\rDo$ since at $z=0$ the argument in the square root becomes $(2-\ODo)^2$ after using the sum rule $\Omo+\ODo=1$.
It is interesting to remark at this point that the behavior of the \DC1\ and \DHlin\ models is rather different from their vacuum counterparts in interaction with matter. Let us e.g. focus on the linear vacuum model, i.e. the model (\ref{eq:linH}) when the EoS is $\wD=-1$ and interacting with matter. As it is shown in Ref.\,\cite{GomezSola2015}, the corresponding DE and matter densities are
\begin{eqnarray}\label{eq:rhomaLinear1}
\rho_{\CC}(z)&=&\rLo\,\left\{1+\Omo\left[(1+z)^{3/2}-1\right]\right\}\\
\rho_m(z)&=&\rho_m^0\left[\Omo+(1-\Omo)(1+z)^{-3/2}\right]\,(1+z)^{3}\label{eq:rhomaLinear2}\,.
\label{eq:rhoLaLinear2}
\end{eqnarray}
In the matter density formula we note an extra factor of $\Omo$ in front of the term $(1+z)^3$ for large enough $z$, as it was already pointed out in \cite{GomezSola2015}. Thus, the deviation with respect to the $\CC$CDM behavior becomes increasingly large when we explore deeper our past. This is in contrast to its \DHlin\  counterpart since matter is conserved for this model and hence it stays with the standard law $\rmr(z)=\rmo (1+z)^{3}$. On the other hand, if we compare the dynamical DE densities \eqref{eq:rDzDH} and \eqref{eq:rhomaLinear1}, both deviate with respect to the $\CC$CDM  linearly with $z$ near our time (as can be seen by expanding these expressions for low $z$), but in in the \DHlin\ case the relative deviation is a factor $1-2/(1+\Omo)>50\%$ larger. This fact is actually determinant to explain the inability of the \DHlin-model to correctly describe the low $z$ data, and it reflects in a bad quality fit in Table 1. We conclude that the linear model (\ref{eq:linH}) is essentially inefficient for a correct description of the cosmological data, both as a vacuum model and as a $\mathcal{D}$-model.
A similar situation occurs for the  \DC1\ and C1 models (cf. Table 1 and \cite{GomezSola2015}). This is a clear sign that the ${\cal D}$-models generally depart significantly from their vacuum analogs and both of them also with respect to the $\CC$CDM. In Sect. 5 we will retake this important issue in more detail, as these models have been repeatedly called for in the literature from different points of view and it may be appropriate to further discuss the reason for their  delicate phenomenological status, which is made quite evident from the statistical analysis presented in the last four columns of Table 1. As we will see, it is not only caused by their background cosmological behavior but also by their troublesome prediction concerning structure formation.

Let us now address the EoS analysis of these models. Using \eqref{eq:EoS} and \eqref{eq:rDforDC1} we are lead to the following expression for the dynamical EoS of the \DC1\ and \DHlin-models:
\be
\label{eq:wEoS}
\omega_D(z)=-1+\frac{\Omega_m^{(0)}(1+z)^3[\epsilon+2\nu E(z)]}{E(z)\,[\epsilon+\nu E(z)]\,[2(1-\nu)E(z)-\epsilon]}\,.
\ee
As we are interested in the behavior of $\omega_D(z)$ and $q(z)$ in the low-redshift period (i.e. near our time), we have neglected the radiation contribution when we compute them. The corresponding EoS plot for \DC1\ and \DHlin\ are shown in Fig.\,2 (right). In the case of \DC1\ ($\nu\neq0$), it displays an asymptote at large $z$ near $2$. The asymptote appears because the fitted value of $\nu$ is negative in Table 1, so the last term of the expression (\ref{eq:rDforDC1}) takes over at sufficiently large $z$ and $\rD$ vanishes near $z=2$ (cf. Fig. 1, right). There is no asymptote for \DHlin\ because the function (\ref{eq:rDzDH}) is monotonically increasing with $z$. Moreover, one can show analytically that for large $z$ the EoS for this model tends to $-1/2$, as can be confirmed graphically by looking at Fig.2.

Of particular importance is the present-day value of the EoS parameter for \DC1\,, i.e. $\omega_D^{(0)}$. Working out the result from the previous equations one finds:
\be
\label{eq:wEoSz0}
\omega_D^{(0)}=-\frac{\epsilon}{(1-\Omega_m^{(0)})(\epsilon+2\Omega_m^{(0)})}\,.
\ee
Notice that this is an exact result and we recall that $\epsilon$ and $\nu$ are \emph{not} small parameters for DC1 models.
Substituting the best-fit values  of the parameters in the above formula, according to the results shown in Table \ref{tableFit2}, the current value of the EoS parameter achieved for $\mathcal{D}$C1  reads $\omega_D^{(0)}=-0.906$\,\footnote{Here and hereafter the numerical estimates use always the barred quantities in Table \ref{tableFit2}, which are the most optimized ones since they are obtained from fitting all the expansion and structure formation data.}. This value can be spotted in Fig.\,2 (right) and deviates significantly from the expected, narrow, range around $-1$\,\,\cite{Planck2015}. This goes, of course, also in detriment of model \DC1\,.
As for the pure linear model \DHlin\, (corresponding to $\nu=0$), the constraint given above enforces $\epsilon=\OLo$ (when we neglect the radiation component) and therefore Eq.\,(\ref{eq:wEoSz0}) yields
\be
\label{eq:wEoSz0H2}
\omega_D^{(0)}=-\frac{1}{1+\Omo}\simeq -0.778
\ee
for the best-fit value of $\Omo$  collected in Table \ref{tableFit2}. Such EoS result is out of the typical range of current observations (even more than for \DC1) and puts the \DHlin\ model once more against the wall.
Let us also note that another particular case of \DC1 models, namely the pure quadratic $\rD\sim H^2$ model (which is obtained for $\epsilon=0$), would have $\wD^{(0)}=0$ according to Eq.\,\eqref{eq:wEoSz0}, and therefore it is completely excluded. The pure quadratic model was used in the past motivated by holographic ideas. But unfortunately the simplest ideas of this sort are actually ruled out\,\cite{MiaoLi04} and some other generalizations (which include \DC1-type models) too\,\cite{BasPolarSola12,BasSola2014b,Japonesos}.

The deceleration parameter for the \DCU-model can be computed from \eqref{eq:qz} and \eqref{eq:HubbleDC1}, and we find

\begin{equation}
\label{qz1}
q=-1+\frac{3\Omega_{m}^{(0)}H_{0}^{2}\left(1+z\right)^{3}}{\left[2\left(1-\nu\right) H^{2}-\epsilon H_{0}H\right]}\,.
\end{equation}

The best-fit values indicated in Table 1 lead to the following estimates for the current acceleration parameters for the general $\mathcal{D}$C1 and \DHlin, respectively: $q^{(0)}=-0.404$, $q^{(0)}=-0.333$. These models are significantly less accelerated now than the $\CC$CDM (for which $q^{(0)}=-0.571$). The behavior of $q(z)$ is plotted in Fig. 3 (right). As it is seen in this figure, for the \DCU-model the Universe has a transit from a decelerated phase ($q > 0$) to an accelerated one ($q < 0$), which occurs precisely at the following transition redshift:
\be
z_{tr}=\left[\frac{2\epsilon^2}{\Omega_m^{(0)}(\epsilon+\Omega_m^{(0)})}\right]^{1/3}-1\,.
\ee
Once more we borrow the fitting results from Table 1 and for $\mathcal{D}$C1 models we obtain $z_{tr}=0.658$, whilst for \DHlin\ we have $z_{tr}=0.528$. These values are clearly smaller than for the $\CC$CDM ($z_{tr}\simeq 0.691$), meaning that the transition is accomplished much more recently than in the standard case.

Let us now finally deal with \DCD-models. This model is sometimes also related with the entropic-force formulations, see \cite{Frampton2010}. However it is not enough to give the structure of the DE to know if the model is phenomenologically allowed or excluded.  Treated as a vacuum C2-model ($\wD=-1$) it was shown in \,\cite{BasSola2014b,GoSolBas2015} to be excluded, but as a ${\cal D}$-model it has some more chances to survive. Let us describe here the main traits of its background evolution and leave the issues of structure formation for the next section. The Hubble function for \DC2\ follows from (\ref{eq:E2DA2}) for $C_0=0$. In the matter-dominated and current epochs reads
\be\label{eq:E2DC2}
E^2(a)=a^{3\beta}+\frac{\Omega_m^{(0)}}{1-\nu+\alpha}(a^{-3}-a^{3\beta})\,.
\ee
As before $\beta\equiv(1-\nu)/\alpha$, but we should recall that in this case $\alpha$ and $\nu$ cannot be both small in absolute value. This is confirmed by  explicitly displaying the constraint $C_0=0$ that they satisfy in the matter-dominated epoch, namely $1-\Omo-\nu+\alpha=-\alpha\wDo\ODo$, in which $\wDo\simeq -1$.  The evolution of the DE density for these models simply follows from (\ref{eq:rDaDA2}) by setting $C_0=0$, and we find:
\begin{equation}\label{eq:rDaDC2}
\rD(z)=\rco\frac{1-\nu+\alpha-\Omo}{1-\nu+\alpha}\,(1+z)^{-3\beta}+\rco\,\Omo\frac{\nu-\alpha}{1-\nu+\alpha}\,(1+z)^{3}\,.
\end{equation}
As can be checked it satisfies $\rD(0)=\rco (1-\Omo)=\rDo$ and it identically reduces to $\rDo$ for $\nu=\alpha=1$, which is the $\CC$CDM result. Another way to see it is that, in this same limit, we have $\beta\to 0$ and the normalized Hubble function (\ref{eq:E2DC2}) becomes exactly the $\CC$CDM one. Ultimately the reason for this is the following: for a DE density of the form \DC2\ in \eqref{eq:ModelsC}, the second Friedmann Eq.\,(\ref{eq:PressureEq}) can be satisfied identically if we set $\omega_D=-1$ and $\alpha=\beta=1$, and at the same time neglect the radiation component. Needless to say, the self-conserved DE equation (\ref{eq:DEConsEq}) is also automatically satisfied, because it is not independent of the two Friedmann's equations. None of the other models could fulfil these conditions and consequently a nontrivial EoS different from $\wD=-1$ is required for them. It is thus not surprising that the fitting procedure singles out a parameter region for \DC2\ very close to the $\CC$CDM (cf. Table 1). In actual fact the fitting values of the parameters do carry some small deviations from unity, as this helps to slightly improve the quality of the fit as compared to the strict $\CC$CDM. This in turn induces a nontrivial evolution of the EoS, which eventually departs from $\wD=-1$ when we move to higher redshifts, so in practice \DC2\ mimics the $\CC$CDM only near our time and quickly deviates from it for $z\gtrsim1$ (cf. Fig.\,2, right).

In contrast to its vacuum counterpart -- viz. the C2 model studied in \cite{BasSola2014b,GoSolBas2015} -- the \DC2-model does have a transition point from deceleration to acceleration.  A straightforward calculation yields
\be
z_{tr}=\left[-\omega_D^{(0)}\frac{\Omega_D^{(0)}}{\Omega_m^{(0)}}(3-3\nu+2\alpha)\right]^{\frac{1}{3(1+\beta)}}-1
=\left[\frac{(1-\nu+\alpha-\Omo)(3-3\nu+2\alpha)}{\alpha\Omega_m^{(0)}}\right]^{\frac{\alpha}{3(1-\nu+\alpha)}}-1\,.
\ee
As expected, for $\nu=\alpha=1$ (which implies $\wDo=-1$ from the aforementioned constraint) we recover exactly the $\CC$CDM result (\ref{eq:ztrLCDM}). As for the effective EoS of this model, $\wD(z)$, it can be directly computed with the help of \eqref{eq:EoS} and (\ref{eq:rDaDC2}). In particular one can check that for $z=0$ we obtain $\wD(0)=-(1-\Omo-\nu+\alpha)/(\alpha\ODo)=\wDo$, which is consistent with the constraint obeyed by the parameters of this model.  For the numerical analysis presented in Table 1 we have fixed $\wDo=-1$ for this model and we have fitted $\Omo$ and $\nu$, and in this way $\alpha$ becomes determined:
$\alpha=(\nu+\Omo-1)/\Omo$. For this reason we can use $\nu$ as the only vacuum free parameter for \DC2\,, and indeed this is the value that is meant for $\nueff$ in Table 1.
The plots for the densities, the EoS behavior, as well as for the deceleration parameter and the transition point from deceleration to acceleration for \DC2\ are included in Figures 1-3. The model works well at low $z$ but we should warn the reader that it may present serious difficulties in describing the radiation epoch, and in fact this is the reason why we have used only the low $z$ observables in its fit. This fact puts us on guard concerning the eventual viability of \DC2\,. We will come back to this important point in Sect. \ref{sect:Discussion}.

%%%%%%%%%%%%%%%%%%%%%%%%%%%%%%%%%%%%%%%%%%%%%%%%%%%%%%%%%%%%%%%%%
%%%%%%%%%%%%%%%%%%%%%%%%%%%%%%%%%%%%%%%%%%%%%%%%%%%%%%%%%%%%%%%%%
%%%%%%%%%%%%%%%%%%%%%%%%%%%%%%%%%%%%%%%%%%%%%%%%%%%%%%%%%%%%%%%%%
\section{Linear structure formation  with self-conserved DE and matter}
\label{sect:LSF}

In this section we deal with the cosmic perturbations for linear structure formation, and we commence with a study of the set of perturbation equations for a general system of self-conserved dynamical dark energy and matter. Subsequently we specialize our results for the concrete  ${\cal D}$-models whose background behavior has been elucidated in the previous sections, all of them falling in that category.

\subsection{Linear perturbations for matter and dark energy}

Let us consider a general system in which the various components (matter and DE) are covariantly self-conserved and the gravitational coupling G remains constant throughout the cosmic history. In the study of linear structure for this system we take into account both the matter perturbations, $\delta\rho_m$, and the perturbations in the DE component, $\rho_D$. Although the DE is not coupled to the matter sector at the background level owing to the assumption of separate covariant conservation, the two sectors develop some kind of interaction in the linear perturbation regime, which we are going to compute. We start by perturbing the $(0,0)$ component of Einstein's equations and the $(0,i)$ components of the covariant energy conservation equation, i.e. $\nabla_\mu G^{\mu\nu}=0$, using the synchronous gauge, i.e. $ds^2=dt^2+(-a^2\delta_{ij}+h_{ij})d\vec{x}^2$, and taking into account that all the components of the cosmic fluid are covariantly self-conserved. We obtain the following coupled system of diferential equations\,\footnote{See e.g. Refs.\,\cite{GrandePelinsonSola08,GrandePelinsonSola09,GSFS10} for a formulation of the perturbation equations in a system of several components, including the DE. Here we have introduced explicitly the perturbations in the dynamical EoS variable of the DE.}:

\be\label{eq:a}
\dot{\hat{h}}+2H\hat{h}=8\pi G\left[\prm+\prD(1+3\omega_D)+3\rho_D\poD\right]
\ee

\be\label{eq:b}
\rho_m\left(\theta_m-\frac{\hat{h}}{2}\right)+3H\prm+\dot{\prm}=0
\ee

\be\label{eq:c}
\rho_D(1+\omega_D)\left(\theta_D-\frac{\hat{h}}{2}\right)+3H[(1+\omega_D)\prD+\rho_D\poD]+\dot{\prD}=0
\ee

\be\label{eq:d}
\rho_m(\dot{\theta_m}+5H\theta_m)+\theta_m\dot{\rho}_m=0
\ee
\be\label{eq:e}
\left\{\rho_D(1+\omega_D)(\dot{\theta}_D+5H\theta_D)+\theta_D\left[\dot{\rho}_D(1+\omega_D)+\rho_D\dot{\omega}_D\right]\right\}\frac{a^2}{k^2} -(\omega_D\prD+\rho_D\poD)=0\,,
\ee
where $\theta_N\equiv\nabla_\mu\delta U_N^\mu$ is the divergence of the perturbed velocity for each component (matter and DE), $\hat{h}\equiv \frac{d}{dt}\left(\frac{h_{ii}}{a^2}\right)$ and $k$ stands for the wave number (recall that these equations are written in Fourier space\,\cite{GrandePelinsonSola08}). %Furthermore, we have used
%non-adiabatic perturbations for $\rho_D$, i.e
%$\dot{p}_D=\dot{\omega}_D\rho_D+\omega_D\dot{\rho}_D$ and $\delta{p}_D=\rho_D\delta{\omega}_D+\omega_D\delta{\rho}_D$.

\subsection{Reduction to a single third-order differential equation for matter perturbations}
\label{sect:ThirdOrder}

Notice that we have six unknowns and five equations, so at this stage is not possible to solve the system in order to find each of the perturbed functions $\theta_m$, $\theta_D$, $\hat{h}$, $\poD$, $\prD$ and $\prm$.  However, we will now show that under reasonable assumptions and working at sub-horizon scales we can eliminate the perturbations in the DE, i.e. $\delta\rD$, in favor of a single higher order equation for the matter part, $\delta\rmr$. This is characteristic of the coupled systems of matter and DE perturbations for cosmologies with self-conserved dynamical DE and matter. For the particular case $\CC=$const. the obtained third-order equation boils down to the (derivative of the) second order one of the $\CC$CDM, as we will show. Let us proceed step by step. Firstly, we make use of \eqref{eq:MatterConsEq} and \eqref{eq:d} so as to obtain $\theta_m(a)$:
\be\label{eq:theta}
\dot{\theta}_m(a)+2H(a)\theta_m(a)=0\rightarrow \theta_m(a)=\theta_{m}^{(0)} a^{-2}\,.
\ee
It follows that the matter velocity gradient decreases fast with the expansion. We shall adopt the conventional initial condition that $\theta_{m}^{(0)}$ is negligible or zero at present and hence $\theta_m\simeq0$ throughout the evolution. Since the scales relevant to the matter power spectrum remain always well below the horizon, i.e. $k\gg H$, we expect small DE perturbations at any time, i.e. the DE should naturally be smoother than matter, wherefore the velocity gradient of the DE perturbations is also naturally set to $\theta_D=0$. In this way we can confine our study to the perturbation $\delta_D$ in the DE sector and $\delta_m$ in the matter sector. This simplification looks reasonable and it will allow us to solve the coupled system of matter and DE perturbations without introducing further assumptions and/or additional parameters.
Under this setup it is clear that the terms that are not proportional to $k^2$ in \eqref{eq:e} can be neglected and we find $\omega_D\prD+\rho_D\poD=0$, which is tantamount to say that we are keeping the perturbations in the DE density but neglecting the perturbations in its pressure, similar as for matter.
From \eqref{eq:b} we can isolate $\hat{h}$
\be\label{eq:h}
\hat{h}=\frac{2}{\rho_m}\left(3H\prm+\dot{\prm}\right)\,,
\ee
and differentiating this expression we arrive at:
\be\label{eq:hdot}
\dot{\hat{h}}=\frac{2}{\rho_m}\left(3\dot{H}\prm+3H\dot{\prm}+\ddot{\prm}\right)-\frac{2\dot{\rho}_m}{\rho^2_m}\left(3H\prm+\dot{\prm}\right)\,.
\ee
Now we insert the last two equations in \eqref{eq:a} and find:
\be\label{eq:at}
B(\prm,\dot{\prm},\ddot{\prm},H,\dot{H})=8\pi G\left[\prm+\prD(1+3\omega_D)+3\rho_D\poD\right]\,,
\ee
where we have defined
\be\label{eq:defB}
B(\prm,\dot{\prm},\ddot{\prm},H,\dot{H})\equiv\frac{6\dot{H}\prm+6H\dot{\prm}+2\ddot{\prm}}{\rho_m}
-\frac{\dot{\rho}_m(6H\prm+2\dot{\prm})}{\rho^2_m}+\frac{4H(3H\prm+\dot{\prm})}{\rho_m}\,.
\ee
On the other hand from \eqref{eq:h} and \eqref{eq:c} we gather
\be\label{eq:ct}
-\frac{\rho_D}{\rho_m}(1+\omega_D)(3H\prm+\dot{\prm})+3H[(1+\omega_D)\prD+\rho_D\poD]+\dot{\prD}=0\,.
\ee
We have three independent equations for the three unknowns: $\poD$, $\prD$ and $\prm$. We first take $\poD=-\omega_D\delta\rho_D/\rho_D$ from the above mentioned equation and substitute it in \eqref{eq:at} and \eqref{eq:ct}. We find:
\be\label{eq:att}
B(\prm,\dot{\prm},\ddot{\prm},H,\dot{H})=8\pi G\left(\prD+\prm\right)
\ee
and
\be\label{eq:ctt}
-\frac{\rho_D}{\rho_m}(1+\omega_D)(3H\prm+\dot{\prm})+3H\,\prD+\dot{\prD}=0\,.
\ee
At this point we can use the last two equations to get rid of the explicit $\prD$ dependence on the DE perturbations, and we obtain:
\be\label{eq:cttt}
-\frac{\rho_D}{\rho_m}(1+\omega_D)(3H\prm+\dot{\prm})+3H\left(\frac{B}{8\pi G}-\prm\right)+\frac{\dot{B}}{8\pi G}-\dot{\prm}=0\,.
\ee
This is the preliminary expression of the final differential equation for the linear density perturbations of the matter field, but in order to make it operative we must still do some algebra. Let us first rewrite the expression $B$, Eq.\,(\ref{eq:defB}), as follows:
\be\label{eq:B}
B=\frac{\prm}{\rho_m}(6\dot{H}+30H^2)+\frac{16H}{\rho_m}\dot{\prm}+2\frac{\ddot{\prm}}{\rho_m}\,,
\ee
and compute its time derivative,
\be\label{eq:Bdot}
\dot{B}=\frac{\prm}{\rho_m}(6\ddot{H}+78H\dot{H}+90H^3)+
\frac{\dot{\prm}}{\rho_m}(78H^2+22\dot{H})+22H\frac{\ddot{\prm}}{\rho_m}+\frac{2}{\rho_m}\dddot{\prm}\,,
\ee
where in the previous formulas we have made repeated use of the matter conservation \eqref{eq:MatterConsEq}.
Introducing the last two expressions in \eqref{eq:cttt}, we obtain the first explicit form of the sought-for third order equation for $\delta\rho_m$:
$$
\frac{\prm}{\rho_m}\left[\frac{6\ddot{H}+96\dot{H}H+180H^3}{8\pi G}-3H\left[\rho_m+\rho_D(1+\omega_D)\right])\right]+\frac{2\dddot{\prm}}{8\pi G\rho_m}+\frac{28H\ddot{\prm}}{8\pi G\rho_m}+
$$
\be\label{eq:inter}
+\frac{\dot{\prm}}{\rho_m}\left[\frac{126H^2+22\dot{H}}{8\pi G}-\left[(\rho_m+\rho_D (1+\omega_D)\right]\right]=0\,.
\ee
A final simplification of Eq.\,\eqref{eq:inter} can still be performed. From the pair of Friedmann's equations \eqref{eq:FriedmannEq} and \eqref{eq:PressureEq} we find the relation
\be
\dot{H}=-4\pi G[\rho_m+\rho_D(1+\omega_D)]\,.
\ee
Using it in \eqref{eq:inter} we finally arrive at the desired form of the third order differential equation for the perturbations of the matter field:
\be\label{eq:inter1}
 \dddot{\prm}+14H\ddot{\prm}+3\dot{\prm}(4\dot{H}+21H^2)+3\prm(\ddot{H}+30H^3+17\dot{H}H)=0\,.\ee
It is convenient to reexpress it in terms the density contrast $\delta_m\equiv \prm/\rho_m$. After some algebra we find the following rather compact form:
%
%$$\dot{\prm}=\dot{\delta}_m\rho_m+\delta_m\dot{\rho}_m$$
%$$\ddot{\prm}=2\dot{\delta}_m\dot{\rho}_m+\delta_m\ddot{\rho}_m+\ddot{\delta}_m\rho_m$$
%$$\dddot{\prm}=3\dot{\rho}_m\ddot{\delta}_m+\dddot{\delta}_m\rho_m+\delta_m\dddot{\rho}_m+3\ddot{\rho}_m\dot{\delta}_m$$
%
\be\label{eq:DifEqCosmicTime}
%\boxed{\dddot{\delta}_m+5H\ddot{\delta}_m+3\dot{\delta}_m(\dot{H}+2H^2)=0}
\dddot{\delta}_m+5H\ddot{\delta}_m+3\dot{\delta}_m(\dot{H}+2H^2)=0\,.
\ee
This is actually the final form in terms of the cosmic time.
However, to make contact with the observations it is highly advisable to rewrite the previous equation in terms of the scale factor, as it has a simple relation with the cosmic redshift. To this end we have to trade the time derivatives for the scale factor derivatives (indicating the latter by a prime), starting from $\dot{\delta}_m=aH\delta_m^\prime$ and its subsequent second and third order derivatives. After a simple calculation we obtain:
%
%$$\dot{\delta}_m=aH\delta_m^\prime$$
%$$\ddot{\delta}_m=a^2H^2\delta_m^{\prime\prime}+(aH^2+a^2HH^\prime)\delta_m^\prime$$
%$$\dddot{\delta}_m=a^3H^3\delta_m^{\prime\prime\prime}+3\delta_m^{\prime\prime}(a^2H^3+a^3H^2H^\prime)+\delta_m^\prime(aH^3+4a^2H^2H^\prime+a^3HH^{\prime %2}+a^3H^2H^{\prime\prime})$$
%
\be\label{eq:DifEqScaleFactor}
%\boxed{\delta_m^{\prime\prime\prime}+\delta_m^{\prime\prime}\left(\frac{8}{a}+\frac{3H^\prime}{H}\right)+\delta_m^\prime\left(\frac{12}{a^2}+\frac{12H^{\prime}}{aH}+\frac{H^{\prime 2}}{H^2}+\frac{H^{\prime\prime}}{H}\right)=0}
\delta_m^{\prime\prime\prime}+\delta_m^{\prime\prime}\left(\frac{8}{a}+\frac{3H^\prime}{H}\right)+\delta_m^\prime\left(\frac{12}{a^2}+\frac{12H^{\prime}}{aH}+\frac{H^{\prime 2}}{H^2}+\frac{H^{\prime\prime}}{H}\right)=0\,.
\ee
In general, this differential equation cannot be solved analytically, but we can proceed numerically. In any case we need to set up the initial conditions at a high redshift, when dust dominates over the dark energy i.e. typically at $z={\cal O}(100)$. The initial conditions are, of course, model-dependent, but it can be shown that for  \DA-models they reduce to the situation that comprises the well-known standard $\Lambda$CDM ones for the function and the first derivative, viz. $\delta_m(a_i)=a_i$ and $\delta_m^\prime(a_i)=1$, to which we have to add $\delta_m^{\prime\prime}(a_i)=0$.

Let us recall that the obtained third order differential equation for the matter perturbations does automatically incorporate the leading effect of the DE perturbations and is valid under the assumption that the values of the perturbed matter and DE velocity gradients are negligibly small. This is the simplest assumption we can make in order to solve the initial system of differential equations without introducing additional parameters; and it is certainly well justified if we are interested in the physics at scales deep inside the Hubble radius (sub-horizon scales), i.e $k\gg H$. From Eq.\,(\ref{eq:e}) we can see that deviations from this framework should scale roughly as $\sim H^2/k^2$. We will estimate them in the next section \ref{sect:DEorNot}.

\subsection{Recovering the \texorpdfstring{$\CC$}{Lambda}CDM perturbations as a particular case}

Let us consider the situation of the $\CC$CDM model deep in the matter-dominated epoch when we can neglect the $\CC$-term. In this case $H^2\propto a^{-3}$ and hence $H^{\prime}(a)/H(a)=-3/(2a)$ and $H^{\prime\prime}(a)/H(a)=15/(4a^2)$. It follows that Eq.\,\eqref{eq:DifEqScaleFactor} boils down in this case to the very simple form
\begin{equation}\label{eq:ReducedForm}
{\delta}_m^{\prime\prime\prime}+\frac72\,\frac{{\delta}_m^{\prime\prime}}{a}=0\,,
\end{equation}
whose elementary solution is $\delta_m=c_1\,a+c_2+c_3\,a^{-3/2}$. Neglecting the decaying mode and setting $c_2=0$ we are led to the standard growing mode solution of the $\CC$CDM, i.e. $\delta\propto a$. This is a clear indication that the $\CC$CDM result is contained in the generalized perturbations equation \eqref{eq:DifEqScaleFactor}.

Alternatively we can prove in a more formal way, in this case using the cosmic time variable, that the well-known $\Lambda$CDM linear perturbation equation\,\cite{PeeblesBook1993,LiddleLyth2000},
\be\label{eq:LCDMperturbations}
\ddot{\delta}_m+2H\dot{\delta}_m-4\pi G\rho_m\delta_m=0\,,
\ee
or, written only in terms of the Hubble function,
\be\label{eq:DifLCDM0}
\ddot{\delta}_m+2H\dot{\delta}_m+\dot{H}\delta_m=0\,,
\ee
is a particular case of \eqref{eq:DifEqCosmicTime} if we treat the dark energy component as a rigid cosmological term with the vacuum constant EoS ($\omega_D=-1$). First, we perform the time derivative of the last equation in order to have a third order one:
\be\label{eq:DifLCDM}
\dddot{\delta}_m+2H\ddot{\delta}_m+3\dot{H}\dot{\delta}_m+\ddot{H}\delta_m=0\,.
\ee
To prove that \eqref{eq:DifEqCosmicTime} and \eqref{eq:DifLCDM} are the same when the DE is the traditional $\CC$-term, let us compute its difference. We find:
\be\label{eq:defDelta}
3H\ddot{\delta}_m+6H^2\dot{\delta}_m-\ddot{H}\delta_m=3H\left(\ddot{\delta}_m+2H\dot{\delta}_m-\frac{\ddot{H}}{3H}\delta_m\right)\,.
\ee
Upon differentiating the pressure equation (\ref{eq:PressureEq}) with respect to the cosmic time under the $\CC$CDM conditions (viz. $\dot{\rho}_{D}=0$) we find $\ddot{H}+3H\dot{H}=0$, so the difference \eqref{eq:defDelta} actually reads
\be 3H\left(\ddot{\delta}_m+2H\dot{\delta}_m+\dot{H}\delta_m\right)=0\,,\ee
where in the last step \eqref{eq:DifLCDM0}  has been used. This obviously proves our contention. At the same time it becomes clear from the proof that if the DE was dynamical (with $\dot{\omega}_D\neq 0$) we could not have obtained the previous result, which means that the third-order differential equation \eqref{eq:DifEqCosmicTime} is indeed more general than the standard one \eqref{eq:LCDMperturbations} and covers the entire class of dynamical DE models with separate (local and covariant) conservation of matter and dark energy densities, at fixed gravitational coupling $G$.

\subsection{Running  \texorpdfstring{$\CC$}{Lambda} and running \texorpdfstring{$G$}{G} with matter conservation}

There is another interesting situation that we would like to mention in passing here in which a third order perturbations equation of the same sort is also needed, but under different conditions. It was first encountered in Ref.\,\cite{GSFS10} and is characterized by dynamical vacuum energy  (hence $\wD=-1$) and conserved matter. This case is, in principle, different from the class of models covered in the previous section because the EoS is of the vacuum type. However, it differs also in a second aspect, namely in the fact that  the gravitational coupling $G$ is running or time-evolving along with the vacuum energy density. In this way the Bianchi identity can be satisfied since there is a dynamical interplay between the vacuum and $G$ that permits the conservation of matter -- see \,\cite{GSBP11,GSFS10} for details. We will show now that the cosmic perturbations of the matter field for this system are also governed  by Eq.\,\eqref{eq:DifEqCosmicTime}, or equivalently by \eqref{eq:DifEqScaleFactor}. In Ref.\,\,\cite{GSFS10} it was shown that the matter perturbations follow the third order equation:
\be\label{eq:GPertEq}
{\delta}_m^{\prime\prime\prime}+\frac{{\delta}_m^{\prime\prime}}{2a}(16-9\Omega_m)+\frac{3\delta^\prime_m}{2a^2}(8-a\Omega^\prime_m+3\Omega_m^2-11\Omega_m)=0\,,
\ee
in which  $\Omega_m(a)={8\pi\,G(a)}\rho_m(a)/{3H^2(a)}$.
Equation \eqref{eq:GPertEq} also includes the effects of the perturbations in $\rho_\Lambda$, which have been eliminated in favor of the matter perturbations following a similar procedure as described in the previous section, except that in this case the dynamics of $G$ enters explicitly too. Taking into account that $\dot{H}=aH(a)H^\prime(a)=-4\pi G(a)\rho_m(a)$  it follows that $\Omega_m(a)=(-2/3)a\, H^{\prime}(a)/H(a)$; and using now this relation in \eqref{eq:GPertEq} we immediately recover \eqref{eq:DifEqScaleFactor}. This confirms that \eqref{eq:DifEqScaleFactor} is, in fact, also valid for models with self-conserved matter and a dynamical vacuum triggered by the time variation of G. We conjecture that \eqref{eq:DifEqScaleFactor} may also be appropriate for more general models with self-conserved matter and different possibilities for
the evolutions of $G$ and the DE, but we will not further pursue the framework of $G$-variable models here -- see \cite{SolaGomezCruz2015} for a recent analysis.

\section{Fitting the  \texorpdfstring{$\mathcal{D}$}{DA}A,  \texorpdfstring{$\mathcal{D}$}{DC}C and  \texorpdfstring{$\mathcal{D}$}{DH}H models to the observational data}\label{sect:Fitting}

Up to this point we have carried out a thorough study of both the background equations as well as of the linear perturbation equations that are obeyed by the class of models under consideration. Therefore we are in position to understand the fitting results collected from the beginning in Table 1 and that were used throughout to evaluate a number of observables associated with the background cosmology in Figs.\,1-3. At the same time we need to further elaborate on the implications of these dynamical DE models on the structure formation. But let us first summarize the methodology employed in our statistical analysis of the cosmological data.

\subsection{Global fit observables}

To carry out the fit we have used the expansion history  (SNIa+BAO$_A$+BAO$_{d_z}$) and the CMB shift-parameter, as well as the linear growth data, through the following joint likelihood function:
\be
\mathcal{L}_{tot}(\vec{p})=\mathcal{L}_{SNIa}\times\mathcal{L}_{BAO}\times\mathcal{L}_{CMB}\times\mathcal{L}_{f\sigma_8}\,,
\ee
in which $\vec{p}=(\Omega^{(0)}_m,\nu)$. By maximizing the total likelihood or, equivalently, by minimizing the joint $\chi^2_{tot}$ function with respect to the elements of $\vec{p}$:
\be
\mathcal{\chi}^2_{tot}(\vec{p})=\mathcal{\chi}^2_{SNIa}\times\mathcal{\chi}^2_{BAO}\times\mathcal{\chi}^2_{CMB}\times\mathcal{\chi}^2_{f\sigma_8},
\ee
we obtain the best-fit values for the various parameters. Concerning the data input, we have used the Union 2.1 set of 580 type Ia supernovae of Suzuki et al. \cite{Suzuki2012}. The corresponding
$\chi^{2}_{\rm SNIa}$ function, to be minimized, is:
\begin{equation}\label{xi2SNIa}
\chi^{2}_{\rm SNIa}({\bf p})=\sum_{i=1}^{580} \left[ \frac{ {\cal
\mu}_{\rm th} (z_{i},{\bf p})-{\cal \mu}_{\rm obs}(z_{i}) }
{\sigma_{i}} \right]^{2} \;,
\end{equation}
where $z_{i}$ is the observed redshift for each data point. The
fitted quantity ${\cal \mu}$ is the distance modulus, defined as
${\cal \mu}=m-M=5\log{d_{L}}+25$, in which $d_{L}(z,{\bf p})$ is the
luminosity distance:
\begin{equation}\label{LumDist}
d_{L}(z,{\bf p})={c}{(1+z)} \int_{0}^{z} \frac{{\rm d}z'}{H(z')}\;,
%d_{L}(z,{\bf p})=\frac{c}(1+z) \int_{0}^{z} \frac{{\rm
%d}x}{H(x)} \;,
\end{equation}
with $c$ the speed of light (now included explicitly in some of these formula for better clarity). In
equation (\ref{xi2SNIa}), the theoretically calculated distance
modulus $\mu_{\rm th}$ for each point follows from using
(\ref{LumDist}), in which the Hubble function is the one
corresponding to each model, see Sect.\,\ref{sec:background}, and we have fixed $H_0=70$ km/s/Mpc following the setting used in the Union 2.1 sample. Finally,
$\mu_{\rm obs}(z_{i})$ and $\sigma_i$ stand for the measured
distance modulus and the corresponding $1\sigma$ uncertainty for
each SNIa data point, respectively. The previous formula
(\ref{LumDist}) for the luminosity distance applies only for
spatially flat universes, which we are assuming throughout.

We have also made use of the BAO estimators $d_{z}(z)$ and $A(z)$ collected by Blake et al. in \cite{Blake2011BAO}. The first one can be computed as follows:

\be
d_z(z_i,\vec{p})=\frac{r_s(z_d)}{D_V(z_i)}\,.
\ee
Here
\be
r_s(z_d)=\int_{z_d}^{\infty}\frac{c\,dz}{H(z)\sqrt{3\left(1+\frac{\delta\rho_b}{\delta\rho_\gamma}\right)}}
\ee
is the comoving sound horizon prior to the drag redshift
epoch, $z_d$, namely the epoch at which baryons are released from
the Compton drag of photons and the gravitational stability of the former can no longer be avoided by the photon pressure; and
\begin{equation}
D_{\rm V}(z)\equiv \left[ (1+z)^{2} D_{A}^{2}(z)
\frac{cz}{H(z)}\right]^{1/3}\;,
\end{equation}
is the ``dilation scale'' introduced by Eisenstein\,\cite{Eisenstein2005}.
In this formula, $D_{A}(z)=(1+z)^{-2} d_{L}(z,{\bf p}) $ is the angular
diameter distance. The other useful BAO estimator is the acoustic parameter, $A(z)$, also introduced by Eisenstein as follows\,\cite{Eisenstein2005}:
\begin{equation}\label{defBAOA}
A({z_i,\bf p})=\frac{\sqrt{\Omo}}{E^{1/3}(z_{i})}
\left[\frac{1}{z_i}\int_{0}^{z_i} \frac{dz}{E(z)} \right]^{2/3}\,,
\end{equation}
where $z_i$ is the redshift at which this observable is measured. In all the theoretical expressions not related with the SNIa distance modulus we have used the current value of the Hubble function given by the Planck Collaboration\,\cite{Planck2015}, i.e. $H_0=67.8$ km/s/Mpc. The corresponding $\chi^{2}$-functions for BAO analysis are defined as:
\begin{equation}\label{BAOd}
\chi^{2}_{\rm BAO_{dz}}({\bf p})=\sum_{i=1}^{6} \left[ \frac{ d_{z,\rm th} (z_{i},{\bf p})-d_{z,\rm obs}(z_{i})}
{\sigma_{z,i}} \right]^{2}
\end{equation}
and
\begin{equation}\label{BAOA}
\chi^{2}_{\rm BAO_{A}}({\bf p})=\sum_{i=1}^{6} \left[ \frac{ A_{\rm th} (z_{i},{\bf p})-A_{\rm obs}(z_{i})}
{\sigma_{A,i}} \right]^{2} \;,
\end{equation}
where $z_{i}$, $d_{z,\rm obs}$, $\sigma_{z,i}$, $A_{\rm obs}$ and $\sigma_{A,i}$
can be found in Table 3 of \cite{Blake2011BAO}.

The CMB shift-parameter, which is
associated with the location of the  first peak $l_1^{TT}$ of the CMB temperature perturbation spectrum is given for spatially flat cosmologies by
\begin{equation}\label{shiftparameter}
R=\sqrt{\Omega_{m}^0}\int_{0}^{z_{*}} \frac{dz}{E(z)}\,,
\end{equation}
where $z_*$ is the redshift of decoupling. The fitted formulae for the baryon drag redshift, $z_d$, and the decoupling redshift,  $z_*$, are given in\,\cite{SugiyamaWu1995,EisensteinWu1998}. The experimental value for $R$ has been taken from \cite{Shaf13}. In this way we have constructed the $\chi^2$ part contributed by the CMB shift parameter as follows:
\be
\chi_{CMB}^2(\vec{p})=\left(\frac{R(\vec{p})-R_{obs}}{\sigma_R}\right)^2\,.
\ee
Finally, we have also taken into account the data on the linear growth rate of clustering provided from different sources, see specifically \cite{Percival2004,Tegmark2006,Guzzo2008,Percival08,Blake2011,Hudson2012,Samushia2012,Beutler2012,Tojeiro2012, Reid2012}. More concretely, we have used $f(z)\sigma_8(z)$ and the corresponding $\chi^2$-function:
\be
\chi_{f\sigma_{8}}^2(\vec{p})=\left(\frac{f\sigma_{8}(\vec{p})-f\sigma_{8obs}}{\sigma_{f\sigma_{8}}}\right)^2\,.
\ee
In recent papers by some of us\,\cite{GoSolBas2015,GomezSola2015} we have elaborated a more detailed explanation of all these cosmological observables as well as on the fitting procedure, and therefore we have left these details aside from the present work. We refer the interested reader to the mentioned references for more information, see also \cite{BPS09,GSBP11}. In the rest of Sect.\,\ref{sect:Fitting} we focus on the observables used to fit the structure formation data, as we feel that they play a very important role in our analysis after we have derived in the previous section the general perturbation equations for the class of self-conserved matter and DE models; and of course we report also on the impact on our models. In particular, we wish to compare the results of our fit analysis in the situation when the DE perturbations are included (as analyzed in the previous section) and when the DE perturbations are not included and the effects of the DE act only on the background history.

\begin{figure}[!t]
\begin{center}
\includegraphics[scale=0.60]{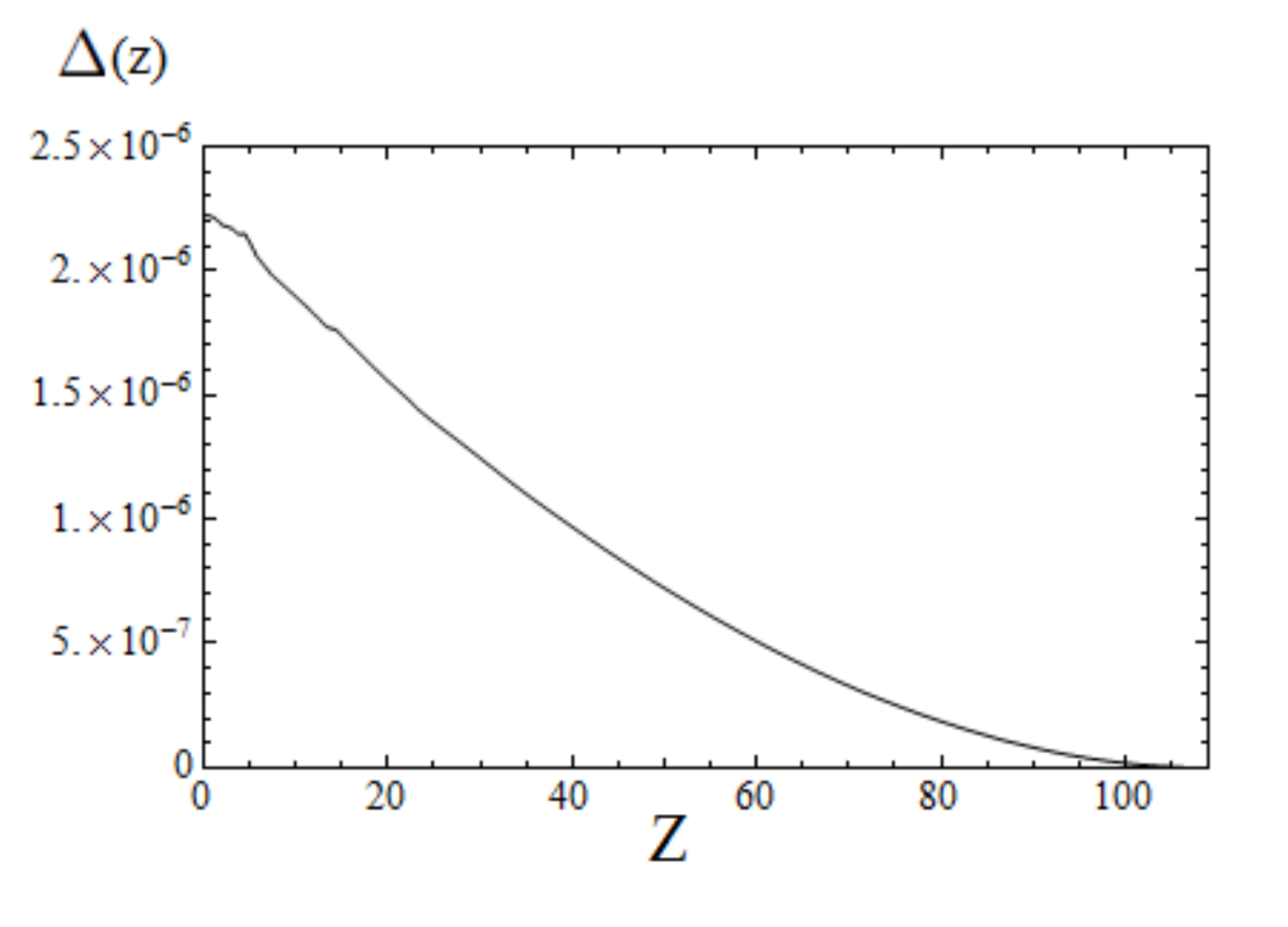}
\caption{\footnotesize{
Relative differences $\Delta(z)$, Eq.\, (\ref{eq:DefDelta}), between the matter density contrasts $\delta_m^{\rm DE}(z)$ computed with the third order differential equation \eqref{eq:DifEqScaleFactor}, which includes DE perturbations, and the result $\delta_m(z)$ from the second order one \eqref{eq:LambdaDifeq}, where the DE density enters only at the background level. The curve shows the essentially degenerate result valid for all the $\mathcal{D}$A models.
\label{fig:DifEqComparison}}
}
\end{center}
\end{figure}
\subsection{Structure formation with and without DE perturbations}\label{sect:DEorNot}

As we have explained in Sect. \ref{sect:ThirdOrder},
we compute the linear matter perturbations $\delta_m$ of the models under consideration through the third-order equation \eqref{eq:DifEqScaleFactor}, which already encodes the effect of the DE perturbations. When the DE perturbations are not included we use the $\CC$CDM form (\ref{eq:LCDMperturbations}), which in terms of the scale factor variable can equivalently be written as
\be\label{eq:LambdaDifeq}
\delta^{\prime\prime}_m+\left(\frac{3}{a}+\frac{H^\prime}{H}\right)\delta^\prime_m+\frac{H^\prime}{aH}\delta_m=0\,,
\ee
with the understanding that the Hubble function in this expression involves the DE density of the corresponding ${\cal D}$-model in Sect\,\ref{sec:DEmodels}. Thus, when using this approximation, the DE effects enter only at the background level. It is surely interesting to compare the results obtained in this approximate way with those generated from the more accurate treatment when the DE perturbations are compute from Eq.\,\eqref{eq:DifEqScaleFactor}. We do not foresee, however, large differences between the two treatments since the DE is expected to be much smoother than matter, at least at the scales where linear structure formation takes place. Still, the fact that we have been able to provide a combined treatment of the DE and matter perturbations in these models gives us a good opportunity to test these expectations.

In Fig.\,4 we provide a quantitative test. We have plotted the relative difference
\be\label{eq:DefDelta}
\Delta (z)=\frac{\delta_m^{\rm DE}(z)-\delta_m(z)}{\delta_m(z)}
\ee
as a function of the redshift. The density contrast $\delta_m^{\rm DE}(z)$ corresponds to the solution of the third order equation \eqref{eq:DifEqScaleFactor} and therefore involves the effect of the DE perturbation, whereas $\delta_m(z)$ is the solution to the more conventional second order equation (\ref{eq:LambdaDifeq}), expressed both in terms of the cosmic redshift $z=(1-a)/a$. As we can see the differences well inside the horizon are very small, of order $\Delta\sim 10^{-6}$. Let us recall that the observational data concerning the linear regime of the matter power spectrum lie in the approximate wave number range $0.01 h {\rm Mpc}^{-1} \lesssim k \lesssim 0.2  h {\rm Mpc}^{-1}$, whereas  the current horizon is given by $H_0^{-1}\simeq 3000\,h^{-1} {\rm Mpc}$ and hence $H_0\simeq 3.33\times 10^{-4} h {\rm Mpc}^{-1}$. From our discussion in Sect. \ref{sect:ThirdOrder} we expect that when we take larger and larger scales (i.e. smaller values of $k$ as compared to $H$) potentially important contributions scaling as $\sim H^2/k^2$ can develop. Clearly the ratio squared of $H_0$ to the minimum and maximum values of the wave number in the linear regime satisfies $H_0^2/k^2\simeq 10^{-6}-10^{-3}\ll 1$  and hence it is natural to expect that the DE perturbations are highly suppressed. Our calculation confirms explicitly this fact. Only if we would approach scales of the order of the horizon such suppression would no longer hold and we could expect sizeable effects from them.

With this nontrivial test we can say that the DE perturbations for the  dynamical ${\cal D}$-models under consideration became fully under control; and in fact we find that at subhorizon scales they do not trigger large deviations from the zeroth order (smooth) DE effects that each model already imprints on the background cosmology. Only at scales comparable to the horizon and for non-negligible velocity gradients the DE perturbations can play a role.

\subsection{Linear growth and growth index}\label{sect:LinearGrowth}

In practice it is convenient to investigate the linear structure formation of our models through the so-called linear growth rate,  namely the logarithmic derivative of the linear density contrast with respect to the scale factor, $f(a)=d\ln\delta_m/d\ln a$\,\cite{PeeblesBook1993}, or equivalently, in terms of the redshift:
\be\label{eq:fz} f(z)=-(1+z)\frac{d\ln\delta_m}{dz}\,.\ee
A related quantity is the $\gamma$-index of matter perturbations\,\cite{PeeblesBook1993,WangSteinhardt98}, defined through $f(z)\simeq \Om(z)^{\gamma(z)}$. For the $\CC$CDM we have $\gamma\simeq 6/11\simeq 0.545$, and one typically expects $\gamma(0)=0.56\pm 0.05$ for $\CC$CDM-like models\,\cite{Pouri2014}. The growth index can obviously be reexpressed as
\be\label{eq:gamma}
\gamma(z)\cong\frac{\ln\,f(z)}{\ln\,\Omega_m(z)}\,.
\ee
By definition $\Om(z)=\rmr(z)/\rc(z)$, and hence for the class of self-conserved models under study it reads
\begin{equation}\label{eq:OmzDef}
\Om(z)=\frac{\rmo\,(1+z)^3}{\rmo\,(1+z)^3+\rD(z)}
\end{equation}
 in which $\rD(z)$ is the DE density of the corresponding ${\cal D}$-model. Obviously $\Om(0)=\rmo/\rco=\Omo$, with $\rco=\rmo+\rDo=3H_0^2/(8\pi G)$ the current critical density. The explicit form for $\rD(z)$ for the models under consideration has been determined in Sect.\ref{sec:background}.

In recent times, however, a more useful quantity is found to be the weighted growth rate of clustering, i.e. $f(z)\sigma_8(z)$. This quantity has the advantage of being independent of the galaxy density bias\,\cite{Percival08}. Here $\sigma_8(z)$ is the rms mass fluctuation amplitude on scales of $R_8=8h^{-1}$ Mpc at redshift $z$. It can be computed as follows -- see e.g. \cite{GoSolBas2015} for details:
\be
\sigma_8(z)=\sigma_{8,\Lambda}\frac{\delta_m(z)}{\delta_m^\Lambda(0)}\left[\frac{\int_{0}^{\infty}k^{2+n_s}T^2(\Omega_m,k)W^2(kR_8)}{k^{2+n_s}T^2(\Omega_m^\Lambda,k)W^2(kR_8)}\right]^{1/2}\,,
\ee
with $n_s=0.968\pm 0.006$ and $\sigma_{8,\Lambda}=0.815\pm 0.009$ (cf. \cite{Planck2015}). The smoothing function
$W(kR)=3({\rm
  sin}kR-kR{\rm cos}kR)/(kR)^{3}$ is the Fourier transform of the following geometric top-hat function with spherical symmetry:  $f_{\rm top\,hat}(r)=3/(4\pi R^3)\,\theta_H(1-r/R)$, where $\theta_H$ is the Heaviside function.
It contains on average a mass $M$ within a comoving radius $R=(3M/
4\pi \rho_m)^{1/3}$. In the above expression $T(\Omega_m^{(0)},k)$ is the BBKS transfer function\,\cite{BBKS86,LiddleLyth2000}. Introducing the dimensionless variable $x=k/k_{eq}$, in which
$k_{eq}=a_{eq}H(a_{eq})$ is the value of the wavenumber at the
equality scale of matter and radiation, we can write the transfer
function as follows:
\be\label{jtf}
T(x) =  \frac{\ln (1+0.171 x)}{0.171\,x}\Big[1+0.284 x + (1.18 x)^2 + \, (0.399 x)^3+(0.490
x)^4\Big]^{-1/4}\,.\ee
Notice that $k_{eq}$ is a model-dependent quantity that must be derived for each of the models we are studying in this paper. For the $\CC$CDM we have the standard result\,\cite{LiddleLyth2000}
\be\label{eq:keq}
(\Lambda CDM):\phantom{XXXXXX} k^{\Lambda}_{eq}=H_0\,\Omo\,\sqrt{\frac{2}{\Omega^0_r}}\,e^{-\Omega_b^{0}-\sqrt{2h}\frac{\Omega_b^{0}}{\Omo}}\,.
 \ee
However, for the \DA\ and \DC1 we have to introduce a correction factor, which is analytically calculable. We find:
\begin{eqnarray}
(\mathcal{D}A):\qquad\qquad
k_{eq}&=&k^{\Lambda}_{eq}\left[\frac{3/2}{4\alpha+3(1-\nu)}+\frac{1/2}{1+\alpha-\nu}\right]^{1/2}\label{eq:keqDA}\\
(\mathcal{D}C1):\qquad\qquad k_{eq}&=&\frac{k^{\Lambda}_{eq}}{\sqrt{1-\nu}}\label{eq:keqDC1}\,.
\end{eqnarray}
As we can see, the correction factor in (\ref{eq:keqDA}) depends separately from $\nu$ and $\alpha$, not just on the difference.
The corresponding formulas for $DA1$ and $DA3$ models is found by setting $\alpha=0$ and $\nu=0$, respectively, in the first expression. Notice that the first formula above reduces to the second for $\alpha=0$, as expected from the fact that at high $z$ the model \DC1 behaves as \DA1. This also explains why the effect of the $\epsilon$-term for \DC1 is negligible in this regime and does not appear in (\ref{eq:keqDC1}). As for the correction factor for model \DC2\ it is derived also from (\ref{eq:keqDA}). Recall that once $\nu$ is fixed from the fit value, the parameter $\alpha$ becomes determined for this model under the conditions discussed at the end of Sect.\, \ref{sect:DCModels}.

 \begin{figure}[!t]
\begin{center}
\includegraphics[scale=0.42]{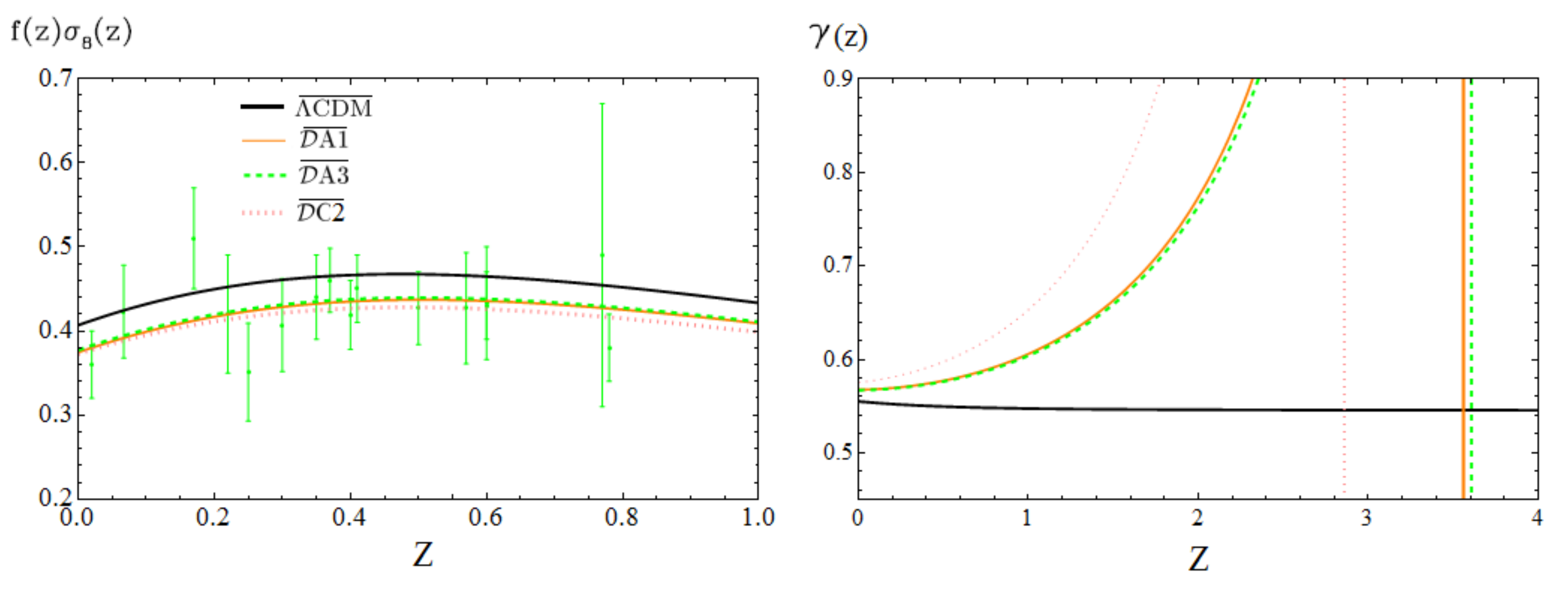}
\caption{\footnotesize{
{\bf Left}: Comparison of the observed and
theoretical evolution of the weighted growth
rate $f(z)\sigma_{8}(z)$ for the $\Lambda$CDM versus the \DA\ and DC2\ models; {\bf Right}: The evolution of the growth index \eqref{eq:gamma} for the same models in a wide span of $z$ so as to display the vertical asymptotes at the points where the DE density  $\rD(z)$ vanishes in each case and the $\gamma$-index (\ref{eq:gamma}) becomes singular. The bar in the models correspond to using the barred fitting parameters of Table 1, whose meaning is explained there.
\label{fig:sigma8gamma}}
}
\end{center}
\end{figure}

The model-dependent corrections to $k_{eq}$ do modify the shape of the transfer function and be of importance in some cases.  It can be instructive for the reader to compare the above correction formulas with the ones determined for the corresponding dynamical vacuum models in \cite{GomezSola2015} -- see Eq.\,(59) in that paper. One may be a bit surprised at first that the formula for the correction factor found there for C1 (denoted in that reference as model II) is considerably more complicated than the one found here for \DC1. The reason is that the matter density for the former model has, at variance with the latter, an anomalous behavior at high redshift  -- cf. our discussion in  Sect.\, \ref{sect:DCModels}, particularly Eq.\,(\ref{eq:rhomaLinear2}). Even so the correction indicated in (\ref{eq:keqDC1}) for \DC1\ can be significant because for this model $\nu$ is not a priori a negligible parameter (cf. Table 1). As a matter of fact, even for models of the \DA-subclass the correction to $k_{eq}$ is of some significance for the overall fit, e.g. it diminishes slightly the $\chi^2$ value with respect to the situation without correction.

\subsection{Results}

Let us now further discuss the specific results obtained for the ${\cal D}$-models under study. Table 1, repeatedly referred to in the previous sections, collects in a nutshell the main numerical output describing the outcome of our analysis. The basic traits of the background history of these models have been presented in Sect.\,\ref{sec:background}. Here we mainly focus on the structure formation results. They are represented in Figs. 5-7. Furthermore, the contour plots combining all the observational data (including of course the structure formation data extracted from the linear growth rate and associated quantities) is indicated in Fig. 8.  Next we discuss these figures in turn.

In Fig.\,5 (left) we consider models \DA\ and \DC2, together with the concordance model $\CC$CDM. We plot their theoretical prediction of $f(z)\sigma_8(z)$ versus the data points (provided in the above mentioned references\,\cite{Percival2004,Tegmark2006,Guzzo2008,Percival08,Blake2011,Hudson2012,Samushia2012,Beutler2012,Tojeiro2012, Reid2012}) as a function of the redshift and for the best fit values of the parameters in Table 1. We use the barred parameters of the table, as in this case we take into account the structure formation data in the fit. Notice that since model \DA2\ interpolates between \DA1\ and \DA3\ we have plotted the last two only in order to show the range that is covered. The fitting results for \DA2\ recorded in Table 1  correspond to fixing $\alpha=-\nu>0$. In fact, since $\alpha$ must be positive (cf. Sect. \ref{sect:BackgroundCosmology}) the previous setting is illustrative of how to break degeneracies among the parameters in a way that is consistent with the sign of $\alpha$. Notice that, at this point, this is necessary since the $\nu$ and $\alpha$ dependence in e.g. equations (\ref{eq:keqDA}) and (\ref{eq:keqDC1}) is no longer of the form $\nu-\alpha$. Other choices compatible with $\alpha>0$ give very similar results.

The reason why we have included \DC2\ also in Fig.\,5  stems from our discussion in Sect.\,\ref{sect:DCModels}, where we showed that this model can mimic the $\CC$CDM near our time, so it is natural to joint it in the same group of models approaching the concordance cosmology now. We can see that the three curves thrive quite well and stay together slightly below the $\CC$CDM line. They actually provide a better fit than the $\CC$CDM (cf. Table 1). We will further qualify the magnitude of this improvement in the next section. On the other hand, in Fig.\,5 (right) we depict the growth index, $\gamma(z)$, defined above. Here the three models are seen to approach together only when $z\to 0$, i.e. around the current time. While \DA1\ and \DA3 still remain quasi-glued at all points, \DC2\ departs significantly from them at larger redshifts. So the mimicking of the $\CC$CDM by \DC2\ is actually not better than that of the \DA-models. Let us also note the presence of vertical asymptotes, which are connected with the vanishing of the denominator in Eq.\,(\ref{eq:gamma}). These are the points where the DE density $\rD(z)$ becomes zero (confer Fig.\,1), and hence $\Omega_m\to 1$ from  Eq.\,(\ref{eq:OmzDef}). Recall also from Fig.\, 2 that at these same points the EoS parameter of the DE becomes singular and develops a corresponding asymptote.

\begin{figure}[!t]
\begin{center}
\includegraphics[scale=0.55]{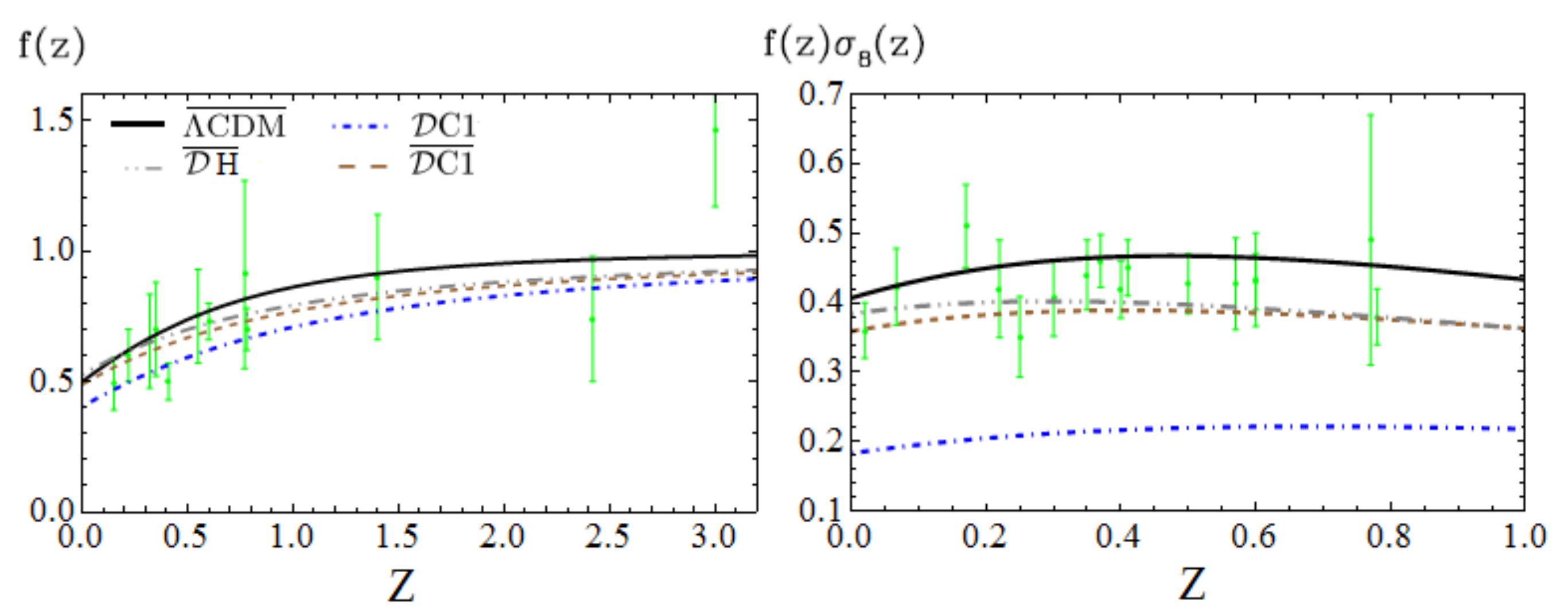}
\caption{\footnotesize{
{\bf Left}: The evolution of the linear growth rate of clustering, Eq.\,(\ref{eq:fz}), specifically for the general \DC1-model and the \DHlin\ one (i.e. the linear model $\rD\sim H$) both for the barred and unbarred fitting parameters indicated in Table 1. For reference, the $\Lambda$CDM model is also included; {\bf Right}: The corresponding curves for the weighted growth rate $f(z)\sigma_{8}(z)$. The dash-dotted line (in blue) in both plots corresponds to the best-fit values of the parameters  for the situation when the structure formation data are not used. This plot makes apparent the delicate situation of the \DC1-model when the unbarred fit parameters are used, particularly in terms of $f(z)\sigma_{8}(z)$ rather than just in terms of $f(z)$ -- see also the text.
\label{fig:sigma8DC1}}
}
\end{center}
\end{figure}

In Fig.\,6  we compare the situation of the \DC1\ and \DHlin\ models (together with the $\CC$CDM as a reference). We have separated these ${\cal D}$-models from the rest because we know that they behave differently. Let us first focus on the plot on the left, where we display the linear growth function $f(z)$ for them, Eq.\,(\ref{eq:fz}). In the case of \DC1\ we include both the curve using the barred parameters and the unbarred ones (cf. Table 1). We can see that in the latter case the \DC1\ curve is displaced only slightly below the others.  However, when we inspect the figure on the right, corresponding to the weighted function $f(z)\sigma_8(z)$, i.e. the growth function appropriately rescaled with the rms mass fluctuation amplitude at each redshift, the same \DC1\ curve now strays downwards quite significantly from the rest. As it turns, the weighted linear growth $f(z)\sigma_8(z)$ appears specially sensitive to that. It demonstrates that when we attempt to fit \DC1\ without using the structure formation data the model immediately gets in tension with them, quite in contrast with the situation with the \DA\ models.

The anomalous behavior of \DC1\  (and, a fortiori, of \DHlin) is particularly clear from the numbers in Table 1, where the difference between the unbarred and barred values of the fitting parameter (viz. $\nu\simeq -0.64$ and $\bar{\nu}\simeq-0.35$) is abnormally large, in contrast to the other models. Such behavior is also reflected in the large increase of $\chi^2$, recorded in Table 1, for this model when we compare $\chi_r^2$, that is to say, the ``reduced $\chi^2$ value'' (the value which is computed without including
the linear growth $\chi^2_{f\sigma_8}$ contribution)  with the unreduced one -- when the growth data are counted in the $\chi^2$ computation. The latter can be computed either without optimizing the fit to the growth data  (yielding $\chi^2/dof=880.74/600$) or after optimizing it (rendering $\bar{\chi}^2/dof=635.23/600$). In both cases the result is much larger than $\chi_r^2/dof=563.86/584$. Obviously the second result is better than the first, but both are extremely poor since they have to be compared with the respective  $\CC$CDM values,  $\chi^2/dof=584.91/608$ and  $\bar{\chi}^2/dof=584.38/608$.  This is  in contrast to the situation with the \DA\ models, where we can check that the difference between the reduced $\chi_r^2$ value and the corresponding $\chi^2$ values computed with or without optimizing the fit to the growth data is by no means as pronounced as in the \DC1\ and \DHlin\ cases.

We can actually retrace the same features described above if we look at Fig.\,7. This is a magnified view of the growth index plot in Fig. 5 (right) for the local range $0<z<2$, where we have superimposed also the growth index for the general
\DC1-model and the linear model \DHlin\,. These lines are in correspondence with the ones in Fig. 6 and highlight once more the anomalous behavior of the \DC1-model and its exceeding sensitivity to the structure formation data.  The current growth index value $\bar{\gamma}(0)\gtrsim 0.66$ (resp. $\gamma(0)\gtrsim 0.72$) that one can read off Fig.\,7 for \DC1\ when the structure formation data are used (resp. not used) in the fit, is clearly too large to be acceptable. The vertical asymptote at $z\simeq 1.5$ (where $\rD(z)$ vanishes) is much closer here because the best-fit value for the unbarred $\nu$ is much bigger in absolute value than  $\bar{\nu}$. As for $\mathcal{D}$H it has no asymptote because the corresponding $\rD(z)$ never vanishes (cf. Fig.\,1). This model also separates from below the $\CC$CDM line. Models \DA\,, instead, perform quite well and meet sufficiently close the $\CC$CDM horizontal line at $\gamma\simeq 6/11\simeq 0.545$  for $z\to 0$.

\begin{figure}[!t]
\begin{center}
\includegraphics[scale=0.5]{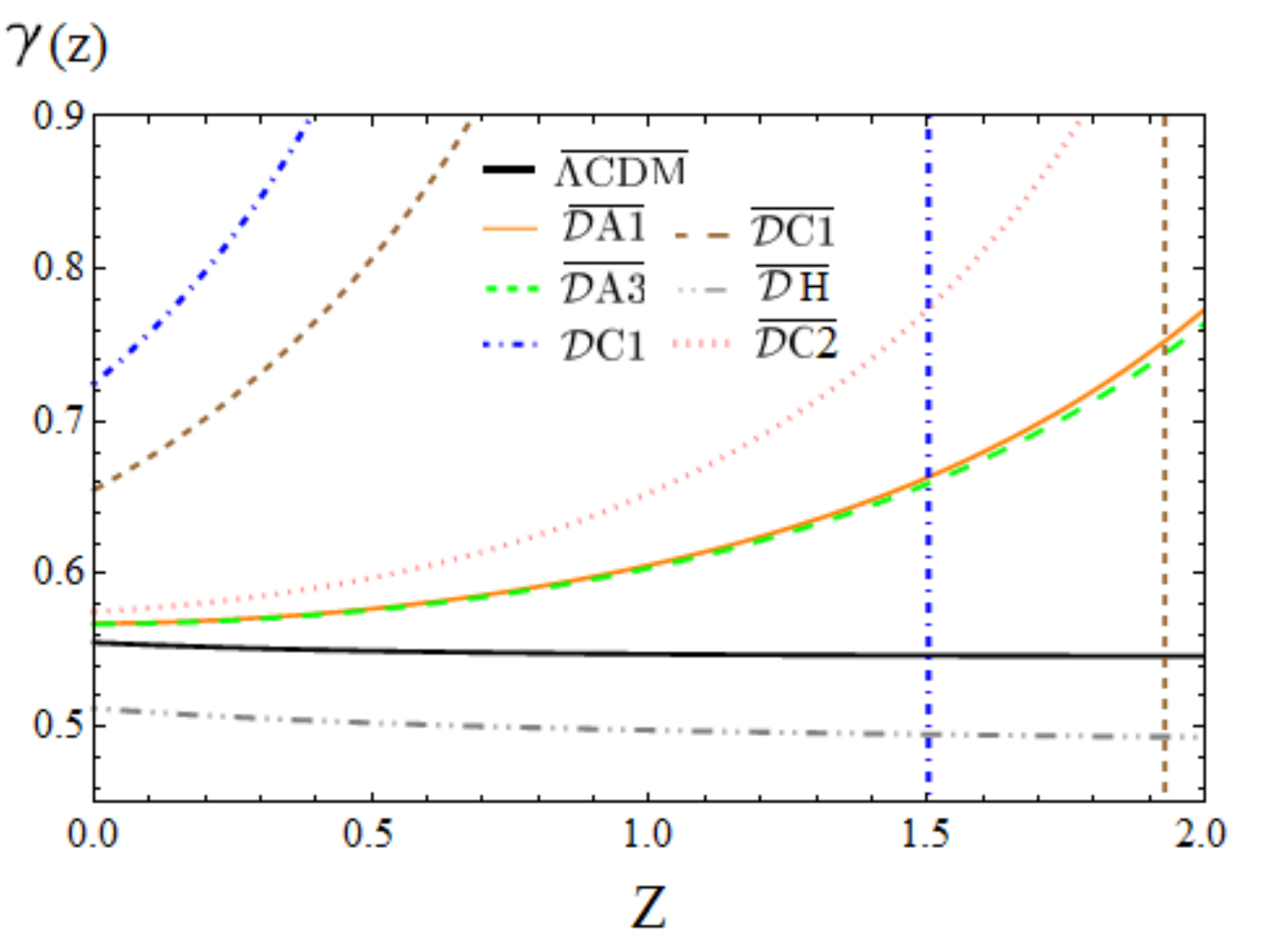}
\caption{\footnotesize{Magnified view of the growth index plot (cf. Fig. 5, right) for the local range $0<z<2$. We have superimposed also the growth index for the general
$\mathcal{D}$C1-model (dash-dotted blue line) and the linear model \DHlin, dash-dotted grey line). The curves have been obtained under the same conditions of the two previous figures.
\label{fig:sigma}}
}
\end{center}
\end{figure}

In short, the unfavorable verdict for \DC1\ seems to have no cure: when we enforce the adjustment of the growth points, then it is the expansion history data that becomes maladjusted; and conversely, if we strive to adjust the expansion history data, then it is the structure formation data that get strayed. In both cases the $\chi^2$ value of \DC1\ becomes too large in comparison to the other models. The exceeding sensitivity to the growth rate and associated quantities is symptomatic of its delicate status as a candidate model to describe the overall observations. And, of course, everything we have said for \DC1\ applies to \DHlin\,, which is the particular case $\nu=0$ of the latter. The model \DHlin\ has no free parameter in the vacuum sector and therefore there is nothing we can do to improve its delicate situation, which in the light of Table 1 appears to be the less favorable one among the models under study.

From the plots in Figs 5-7 and the statistical results of Table 1 we can say that the \DA\ models are capable of describing the overall set of data in a way which is perfectly competitive with the $\CC$CDM. Model \DA1\,, in particular, represents the simplest realization with one parameter, $\nu$. The same can be said of model \DA3\ with the parameter $\alpha$. These models are similar because the time evolutions induced by $H^2$ and $\dot{H}$ are comparable. As for model \DA2\ it is the most general of the \DA-subclass and interpolates between the two previous cases. All of them offer a better fit quality than the $\CC$CDM. In the next section we further qualify this assertion.

%%%%%%%%%%%%%%%%%%%%%%%%%%%%%%%%%%%%%%%%%%%%%%%%%%%%%%%%%%%%%%%%%
%%%%%%%%%%%%%%%%%%%%%%%%%%%%%%%%%%%%%%%%%%%%%%%%%%%%%%%%%%%%%%%%%
%%%%%%%%%%%%%%%%%%%%%%%%%%%%%%%%%%%%%%%%%%%%%%%%%%%%%%%%%%%%%%%%%

\begin{figure}[!t]
\begin{center}
\includegraphics[scale=0.5]{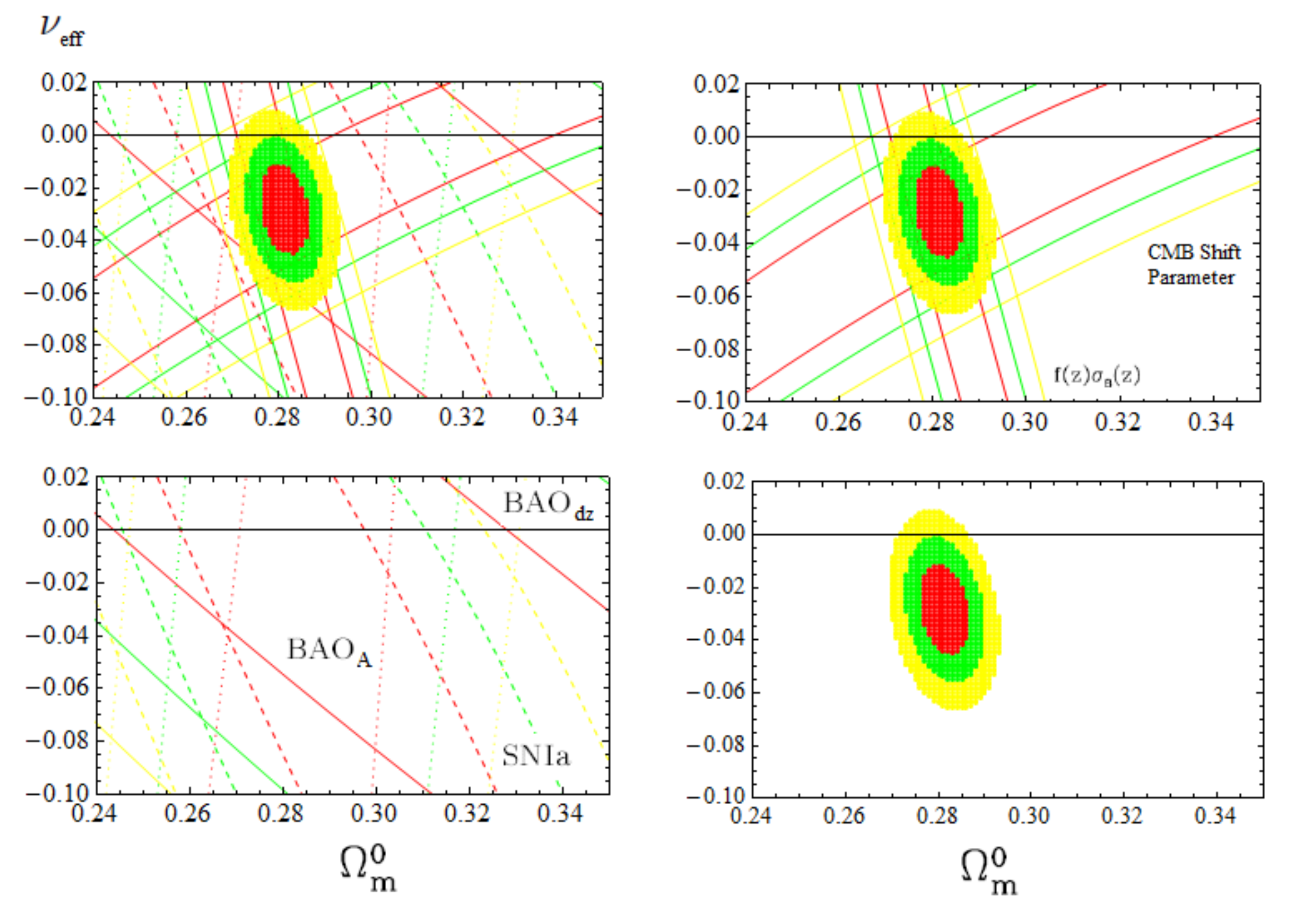}
\caption{\footnotesize{
Likelihood contours in the $(\Omega_m^{(0)},\nueff)$ plane (for the values $-2\ln\mathcal{L}/\mathcal{L}_{max}=2.30$, $6.16, 11.81$, corresponding to 1$\sigma$, 2$\sigma$ and $3\sigma$ confidence levels) for the \DA1-model using the full expansion history and CMB shift parameter data (BAO+SNIa+CMB). Models \DA2\ and \DA3\ present very similar plots, as can be expected from the statistical analysis of Table 1, and for this reason we have denoted $\nueff$ the vertical axis of the plot. In all cases the $\nueff=0$ region ($\CC$CDM) is disfavored at $\sim 3\sigma$ level.
\label{fig:CLW1}}
}
\end{center}
\end{figure}

\subsection{Discussion}\label{sect:Discussion}

Let us now focus on Fig. 8, where we display a detailed representation of the fitting procedure we have followed in order to pin down the most favorable region of the various DE models under study. In that figure we specifically consider the case of model \DA1\,, where we can carefully appraise the way we have combined all the data on expansion history and structure formation mentioned in Table 1. Thanks to the juxtaposition of the permitted regions defined by each observable we find that a well delimited domain emerges. This bounded, elliptically shaped, area is what we may call the physical region for each model. The upper plot on the left of that figure displays all the lines involved in the projection of the final physical region; right next we show a plot with the boundary lines corresponding to the CMB shift parameter and the weighted growth rate $f(z)\sigma_8(z)$; then the lower plot on the left highlights the edges associated with the two types of BAO data involved, together with the borders of the area allowed by the supernovae (SNIa) observations; finally, the net intersection of all of them is represented by the lower plot on the right, in which we depict the $1\sigma$, $2\sigma$ and $3\sigma$ domains within the physical region.
The corresponding plots for models \DA2\ and \DA3\ are very similar and are not shown.

A most remarkable feat emerging from our analysis is the fact that the \DA-subclass of dynamical DE models is capable of improving the fit quality of the $\CC$CDM. This observation is specially significant if we bare in mind that the physical region in Fig.\,8 is mostly located outside the domain of the concordance model, namely it lies roughly $\sim 3\sigma$ away from the $\nueff=0$ line representing the $\CC$CDM; specifically, we can estimate with accuracy that $84.27\%$ of the area of the  $3\sigma$ domain is below the $\nueff=0$ line.
Models \DA\,, therefore, represent a challenge to the $\CC$CDM since they are phenomenologically quite competitive. Somehow they ``break the ice'' inherent to the rigid character of the cosmological $\CC$-term, showing that the possibility of a dynamical vacuum is not only more natural from the theoretical point of view but also more instrumental for adjusting the cosmological observations.

In order to assess the statistical quality of our fits, and the competition score between the models, we have displayed their global $\chi^2$ value per degree of freedom in Table 1. However, in the same table we provide also (in the last columns) the value of another related statistics, which we now describe. It is the so-called Akaike Information Criterion (AIC)\,\cite{Akaike}, which is particularly useful for comparing competing models describing the same data and is amply used in many areas of science --- see e.g. the book\,\cite{IACBook}. We wish to use the AIC here to compare the ${\cal D}$-models with the $\CC$CDM. Such criterion is formulated directly in terms of the maximum of the likelihood function, ${\cal L}$, and is defined as follows. Let $n_p$ be the number of independent estimable parameters in a given model and $N$ the sample size of data points entering the fit. If ${\cal L_{\rm max}}$ is the maximum value of the likelihood function, the corresponding AIC value is defined as follows:
\begin{equation}\label{eq:IACDef}
{\rm AIC}=-2\ln{\cal L_{\rm max}}+2n_{p}+\,\frac{2n_p(n_p+1)}{N-n_p-1}\,.
\end{equation}
The comparison criterion based on this statistics is the following: given two competing models describing the same data, the model that does better is the one with smaller AIC value. Notice that there is a kind of tradeoff between the first term on the r.h.s. of \eqref{eq:IACDef} and the other two. The first term (the one carrying the maximum likelihood value) tends to decrease as more parameters are added to the approximating model (since ${\cal L_{\rm max}}$ becomes larger), while the second and third terms increase with the number of parameters and therefore represent the penalties applied to our modeling when we add more and more parameters. In particular, the second term ($2n_p$) represents an universal penalty related only to the increase in the total number of independent parameters, whereas the third term is an extra penalty applied when the sample is not sufficiently large as compared to the number of independent parameters.
For large samples $N\gg n_p$ (typically for $N/n_p > 40$) the third term becomes negligible and the above formula simplifies to
\begin{equation}\label{eq:IACDef2}
{\rm AIC}=-2\ln{\cal L_{\rm max}}+2n_{p}\,.
\end{equation}
This is actually the situation that holds good in our case. Indeed, for the $\CC$CDM we have one independent parameter ($n_p=1$) represented by $\Omo$, whereas for \DA1\,, for instance, we have $n_p=2$, namely $(\Omo,\nu)$. In the case of \DA2\ we have $(\Omo,\nu,\alpha)$ and hence $n_p=3$, but since we fix a relation between $\nu$ and $\alpha$ we have again $n_p=2$. On the other hand we can see from Table 1 that $N\simeq 590$ in all cases, and therefore the condition $N/n_p > 40$ is amply satisfied and the use of the more simplified formula (\ref{eq:IACDef2}) is fully justified in this case.  For Gaussian errors, the maximum of the likelihood function can be expressed in terms of the minimum of $\chi^2$ and therefore Eq.\,(\ref{eq:IACDef2}) can be reexpressed as
\begin{equation}\label{eq:IACDef3}
{\rm AIC}=\chi^2_{\rm min}+2n_{p}\,.
\end{equation}
In practice, to test the effectiveness of models $M_i$ and $M_j$, one considers the pairwise difference (AIC increment)
$(\Delta$AIC$)_{ij} = ({\rm AIC})_{i} - ({\rm AIC})_{j}$. The larger the
value of $\Delta_{ij}\equiv|\Delta({\rm AIC})_{ij}|$, the higher the evidence against the
model with larger value of ${\rm AIC}$, with $\Delta_{ij} \ge 2$ indicating a positive such evidence and $\Delta_{ij}\ge 6$
denoting significant such evidence. Finally the evidence ratio (ER) against one model whose AIC value is larger than another is judged by the relative likelihood of model pairs, ${\cal L}_i/{\cal L}_j$, or equivalently by the ratio of Akaike weights $w_i=e^{({\rm AIC})_{i}/2}$. Thus, given two models $M_i$ and $M_j$ whose AIC increment is $\Delta_{ij}$, the evidence ratio (ER) against the model whose AIC value is larger is computed from ${\rm ER}=w_i/w_j=e^{{\Delta}_{ij}/2}$.

From Table 1 we can derive the AIC increments $(\Delta$AIC$)_{ij}$ and evidence ratios when we compare the fit quality of models $i=$\DA\,,\DC\,,\DHlin\ with that of  $j=\CC$ ($\CC$CDM).  We find that for all \DA\ models $(\Delta$AIC$)_{i\CC}\gtrsim 9$  against the $\CC$CDM. It follows that when we compare any \DA\ model with the $\CC$CDM the typical evidence ratio against the latter is typically of order ER$\simeq 90$.  Worth noticing is also the result of the fit when we exclude the growth data from the fitting procedure but still add their contribution to the total $\chi^2$ -- it corresponds to the unbarred parameters in Table 1. This fit is of course less optimized, but allows us to risk a prediction for the linear growth and hence to test the level of agreement with these data points. It is noteworthy that in the case of the \DA\ models the corresponding AIC pairwise differences with the $\CC$CDM remain similar to the outcome when the growth data enters the fit. Therefore, the $\CC$CDM appears significantly disfavored versus the \DA-models in both cases (i.e. with barred and unbarred parameters in Table 1).

Quite another story is the situation with the \DC1-model, which does not seem capable to adjust simultaneously the expansion history observables and the structure formation data. A glance at Table 1 shows that when we compare it with the $\CC$CDM, with barred parameters, we find  $(\Delta$AIC$)_{1\CC}\gtrsim 53$  against \DC1\,, whereas with unbarred ones we get the even worse result $(\Delta$AIC$)_{1\CC}\gtrsim 296$ . The respective evidence ratios against \DC1\ versus the $\CC$CDM under these two circumstances are of course extremely large, the smallest one being ER$\sim 10^{11}$. Thus, quite obviously, from the point of view of the Akaike Information Criterion the situation of the \DC1\ model is  desperate, and that of \DHlin\ is even more dramatic. Not so for the \DC2\ model, but we will assess its viability more carefully at the end.

The upshot after making use of the AIC is that the competitive position of the \DA\ models became even more prominent from the point of view of phenomenology, whilst the status of \DC1\ and \DHlin\ against observations appear essentially as terminal. This fact may have nontrivial implications for cosmological model building.  For instance, the \DC1\ model has been studied in some formulations relating QCD with cosmology and it goes sometimes under the name of  ``QCD ghost dark energy model''\,\cite{QCDghost1,QCDghost2,QCDghost3}.  Although some formulations of these models are certainly of theoretical interest and could possibly be modified to become in better harmony with observations, the strict phenomenological analysis of models of DE containing a linear power of the Hubble function and a quadratic power of it, with no additive constant term ($C_0=0$) -- i.e. the strict \DC1-form indicated in our model list (\ref{eq:ModelsC}) --- leads to the inescapable conclusion that they are unable to account for the observations. The particular case in which the DE contains only the linear term in $H$, i.e. $\rD\sim H$, suffers of an even more hapless fate. Such linear form was first proposed in Ref.\,\cite{Schutzhold02} and subsequently  was adopted and adapted to different purposes by different authors, until it eventually gave rise to the extended \DC1\ form. Incidentally, the two \DC\ models (\ref{eq:ModelsC}) and the linear (\ref{eq:linH}) were previously shown to be phenomenologically problematic as well when treated as vacuum models, see \cite{GoSolBas2015} and in particular \cite{GomezSola2015}. Here we have shown that the phenomenological conflict also persists when they are treated as ${\cal D}$-models with self-conserved dynamical DE. If we compare Fig.\,5 of Ref.\,\cite{GoSolBas2015} with the present Fig. 6 we can see that the \DC1-model actually does better than their vacuum counterpart, the C1-model (whose lack of power for structure formation near our time is quite dramatic, as shown in that reference). Still, the overall performance of the \DC1-model remains helpless in front of the $\CC$CDM and its \DA\ companions.

We would like to point out here that our assessment about the \DC1\ and \DHlin\ models is definitely much more pessimistic as compared to the analysis presented e.g. in\,\cite{QCDghost3} -- see also the review \cite{ReviewUniverse}. Take e.g. \DC1\,; the basic difference that we have been able to trace with respect to these authors is that we do not find that such model can pass the structure formation test when \DC1\ is adjusted to the background data, and vice versa. We suspect that in the aforementioned references the linear growth data points have directly been fitted with the \DC1\ model without perhaps verifying if the density perturbations of the model were able to meet the $\CC$CDM behavior at large redshift, namely $\delta\propto a$. In our case we fitted the data after imposing such boundary condition, and we found that the model failed (by far) to reproduce the observations, as we have reported in detail throughout our work. The results exhibited in Figures 6-7, combined with the comparatively poor statistical output of the \DC1-model recorded in Table 1, altogether make a quite eloquent case against the delicate health of that model in front of observations. The situation with \DHlin\,, which is a particular case of \DC1\,, is even worse since it has one less degree of freedom to maneuver. In short, from the present comprehensive study we conclude that the \DC1\ and \DHlin\ models are ruled out. This was already the firmly inferred conclusion when these models were treated within the vacuum class\,\cite{BasSola2014b,GoSolBas2015,GomezSola2015},\footnote{See in particular the thorough analysis of Ref. \,\cite{GomezSola2015}, including the additional details provided in the Appendix of that paper and references therein.} and here we can confirm that it is also the same unfavorable situation in the context of the ${\cal D}$-class.

Let us finally comment on the viability of the \DC2-subclass, which was shown to fit the data reasonably well (cf. Table 1). It is sometimes associated with the so-called entropic-force models\,\cite{Frampton2010}. We have seen in Sect. \ref{sect:DCModels} that it can provide a reasonable fit to the low redshift data (including linear growth) and therefore performs better than the \DC1\ and \DHlin\ models. However, this conclusion cannot be placed out of context, meaning that the structure of the \DC2\ model -- as in fact of any other one in the list of DE models given in (\ref{eq:ModelsA}-\ref{eq:linH}) -- is not sufficient to conclude whether the model is viable or excluded; for it also depends on whether there is exchange of energy with matter or not. Thus, despite the \DC2\ model (as a self-conserved DE model) provides a reasonable fit to the low-$z$ data in Table 1, its vacuum counterpart -- i.e. the model C2 with $\wD=-1$ and matter non-conservation studied in detail in \cite{BasSola2014b,GoSolBas2015} -- was shown to be ruled out, already at the background level, even though it has exactly the same structure as a function of $H$.

Unfortunately \DC2\ has ultimately, too, a very serious drawback as self-conserved DE model, which we have preliminary announced at the end of Sect.\,\ref{sect:DCModels}. While \DC2\ tends to mimic the $\CC$CDM for a while, which is of course a positive attribute, such feature is limited to a period around the current epoch and cannot be extended to the radiation-dominated era. The reason can be seen on inspecting the radiation term in Eq.\,(\ref{eq:E2DA2}), namely the one proportional to $\Oro/(1-\nu+4\alpha/3)$. This term does apply to \DC2\ as well (with $C_0=0$), and the problem appears because the low-$z$ data naturally choose $\nu\simeq\alpha\simeq 1$ (cf. Table 1) so as to mimic as much as possible the $\CC$CDM. However, this is only possible at the price of ``renormalizing'' the size of the radiation coefficient from $\Oro$ to  $\sim 3\Oro/4$. Obviously this is a rather significant modification, in which $25\%$ of the normal content of the radiation is missing. Thus, deep in the radiation-dominated epoch, the model becomes anomalous and one can foresee a significant departure from the $\CC$CDM. For this reason and despite its ostensible success at low energies we do not place this model in the list of our favorite ${\cal D}$-models. A more detailed analysis (which we do not deem necessary here) would require to appropriately modify the standard formulas existing in the literature e.g. for computing the decoupling and baryon drag redshifts, as well as the corresponding modification of the BAO$_{dz}$ observable (cf. the discussion of these formulas in \cite{GoSolBas2015} and references therein).

We close this section by mentioning the analysis of Ref.\,\cite{MathewDC2} on a similar model as our \DC2\,. The authors do not seem to have detected the above mentioned important problem, as in fact they did not compute the radiation contribution. In addition, these authors incorrectly compare their results with those of \cite{BasSola2014b} without realizing, or at least clarifying, that the two models are very different; namely, in one case it is a C1-type vacuum model\,\cite{GoSolBas2015} in interaction with matter, whilst in the other it is a \DC1-model with self-conserved matter and DE. Be as it may, in the light of the results presented here it becomes clear that all the  models with $C_0=0$, whether in the vacuum class or in the ${\cal D}$-class, are actually excluded and cannot be used to support an alternative formulation of holographic DE .

\section{Conclusions}

In this work we have thoroughly analyzed the ${\cal D}$-class of self-conserved dynamical dark energy models. These models are based on a dynamical DE density  $\rD$ which takes the form of a series of powers of the Hubble rate, $H$, and its derivatives. To be consistent with the general covariance of the effective action of quantum field theory in curved spacetime, we can expect that the series of powers must involve an even number of cosmic time derivatives of the scale factor only. Thus, if we should apply strictly this recipe for the current Universe, only the first powers up to $\dot{H}$ and $H^2$ would be involved in the structure of $\rD$, including or not a constant additive term $C_0$. However, in order to cover a wider variety of possibilities that have also been addressed in the literature from different theoretical perspectives, we have admitted also a subclass of models in which the linear term in $H$ is included for $C_0=0$. This greater flexibility in the general form of $\rD$ has allowed us to compare the performance of this restricted set of models with respect to those that are theoretically most preferred, namely the $C_0\neq 0$ ones.

An additional characteristic of the ${\cal D}$-class is that $\rD$ is locally and covariantly self-conserved. At fixed value of the gravitational constant, this entails the simultaneous local covariant conservation of matter as a necessary condition to satisfy the Bianchi identity. The ${\cal D}$-class models are thus enforced to have a dynamical equation of state with a nontrivial evolution with the cosmic expansion, $\wD=\wD(H)$, and therefore it is different from the previously studied dynamical vacuum class\,\cite{GoSolBas2015} in which $\wD=-1$ and the vacuum exchanges energy with matter. The two dynamical DE types share the same formal functional dependence of $\rD$ on $H$, but the behaviors are substantially different, already at the background level, as attested by the analytical and numerical solutions in each case.

We have also studied for the first time (to the best of our knowledge) the generic set of linear matter and DE perturbations for a system of self-conserved cosmic components, and studied the conditions by which the system can be transformed into an equivalent third order differential equation for the matter perturbations. We have demonstrated that the situation of a rigid $\CC$-term, i.e. the $\CC$CDM concordance model, appears as a particular case of that general framework. We have subsequently applied it to the entire ${\cal D}$-class of DE models and upon solving the general perturbation equations we have compared the solution with the corresponding results obtained when the DE enters only at the background level. In this way we have been able to have good control on the reach of the DE perturbations and explicitly confirmed that they play a small role when the considered scales are well under the horizon.

After a detailed study of the background history and the cosmic perturbations of the various models in the ${\cal D}$-class, we have been able to identify the subclass of the \DA-models as a most promising one insofar as it provides an excellent fit to the overall data on Hubble expansion and structure formation. These are the theoretically favored ${\cal D}$-models for which $C_0\neq0$ -- cf. Eq.\,(\ref{eq:ModelsA}). The fit quality rendered by them has been shown to be significantly better than that of the $\CC$CDM. Most conspicuously, using the Akaike Information Criterion\,\cite{Akaike} as a method to compare competing models describing the same data, we find that the evidence ratio in favor of the \DA-subclass (namely the relative likelihood of these models as compared to the concordance model) is of order of a hundred. We also find that the physical region of the parameter space for the \DA-models lies $\sim 3\sigma$ away from the $\CC$CDM region, and therefore the two models can be clearly distinguished.

Using the same testing tools we have reached the firm conclusion that all of the ${\cal D}$-models with $C_0=0$ are strongly disfavored, in particular the linear model $\rD\sim H$. Furthermore, among the theoretical models existing in the literature that become automatically excluded by our analysis we have the so-called entropic-force models and the QCD-ghost formulations of the DE (cf. Sect. \ref{sect:Discussion} for details and references).

At the end of the day the most distinguished dynamical  ${\cal D}$-models, both theoretically and phenomenologically, are those in the \DA-subclass -- viz. the set of models which endow the DE of a mild dynamical character and at the same time have a well-defined $\CC$CDM limit. These models improve significantly the fit quality of the $\CC$CDM, showing that a moderate dynamical DE behavior is better than having a rigid  $\CC$-term for the entire cosmic history.

We have found that the favored \DA-subclass has also the ability to mimic the quintessence behavior and it could even provide a possible explanation for the phantom character of the DE at present, as suggested by the persistent region projected below the $\wD=-1$ line (the so-called phantom divide) from the fits to the recent and past cosmological data. There is of course no significant evidence of this phantom character, as the physical region includes the $\wD=-1$ line. However, if in the future more accurate observations would insist on singling out the domain below the phantom divide, then the dynamical DE models in our \DA-subclass could provide a simple explanation without need of invoking true phantom fields, which are of course abhorred in QFT.

Let us conclude by emphasizing our main message.  The dynamical DE models treat the DE density as a cosmic variable on equal footing to the matter density. In a context of an expanding universe this option may be seen as more reasonable than just postulating an everlasting and rigid cosmological term for the full cosmic history. Furthermore, the subclass of \DA-models favored by our analysis furnishes a theoretically consistent and phenomenologically competitive perspective for describing the dark energy of our Universe as a dynamical quantity evolving with the cosmic expansion. The structure of the DE density in these models as a series of powers of the Hubble rate and its time derivatives is suggestive of a close connection with the QFT formulation in curved spacetime. We hope that they will help to better understand the origin of the cosmological term from a more fundamental perspective, and hopefully they might eventually shed some light as to the nature of the cosmological vacuum energy and its relation with the quantum vacuum of modern gauge field theories.

\vspace{0.5cm}

\acknowledgments The work of AGV has been partially supported by an APIF
grant of the Universitat de Barcelona. EK has been supported in part by Bu-Ali-Sina Univ. Hamedan and is thankful to the Ministry of Science, Research and Technology of Iran for financial support; she would also like to thank the Dept. ECM, Univ. de Barcelona, for support and warm hospitality during the realization of this work. JS has been supported in part
by FPA2013-46570 (MICINN), Consolider grant CSD2007-00042 (CPAN),
2014-SGR-104 (Generalitat de Catalunya) and MDM-2014-0369 (ICCUB).

%%%%%%%%%%%%%%%%%%%%%%%%%%%%%%%%%%%%%%%%%%%%%%%%%%%%%%%%%%%%%%%%%
%%%%%%%%%%%%%%%%%%%%%%%%%%%%%%%%%%%%%%%%%%%%%%%%%%%%%%%%%%%%%%%%%
%%%%%%%%%%%%%%%%%%%%%%%%%%%%%%%%%%%%%%%%%%%%%%%%%%%%%%%%%%%%%%%%%

\newcommand{\CQG}[3]{{ Class. Quant. Grav. } {\bf #1} (#2) {#3}}
\newcommand{\JCAP}[3]{{ JCAP} {\bf#1} (#2)  {#3}}
\newcommand{\APJ}[3]{{ Astrophys. J. } {\bf #1} (#2)  {#3}}
\newcommand{\AMJ}[3]{{ Astronom. J. } {\bf #1} (#2)  {#3}}
\newcommand{\APP}[3]{{ Astropart. Phys. } {\bf #1} (#2)  {#3}}
\newcommand{\AAP}[3]{{ Astron. Astrophys. } {\bf #1} (#2)  {#3}}
\newcommand{\MNRAS}[3]{{ Mon. Not. Roy. Astron. Soc.} {\bf #1} (#2)  {#3}}
\newcommand{\PR}[3]{{ Phys. Rep. } {\bf #1} (#2)  {#3}}
\newcommand{\RMP}[3]{{ Rev. Mod. Phys. } {\bf #1} (#2)  {#3}}
\newcommand{\JPA}[3]{{ J. Phys. A: Math. Theor.} {\bf #1} (#2)  {#3}}
\newcommand{\ProgS}[3]{{ Prog. Theor. Phys. Supp.} {\bf #1} (#2)  {#3}}
\newcommand{\APJS}[3]{{ Astrophys. J. Supl.} {\bf #1} (#2)  {#3}}
%%%%%%%%%%%%%%%%%%%%%%%%%%%%%%%%%%%%%%%%%%%%%%%%%%%%%%%%%%%%%%%%%%%%%%%%%

\newcommand{\Prog}[3]{{ Prog. Theor. Phys.} {\bf #1}  (#2) {#3}}
\newcommand{\IJMPA}[3]{{ Int. J. of Mod. Phys. A} {\bf #1}  {(#2)} {#3}}
\newcommand{\IJMPD}[3]{{ Int. J. of Mod. Phys. D} {\bf #1}  {(#2)} {#3}}
\newcommand{\GRG}[3]{{ Gen. Rel. Grav.} {\bf #1}  {(#2)} {#3}}

%%%%%%%%%%%%%%%%%%%%%%%%%%%%%%%%%%%%%%%%%%%%%%%%%%%%%%%%%%%%%%%%%
%%%%%%%%%%%%%%%%%%%%%%%%%%%%%%%%%%%%%%%%%%%%%%%%%%%%%%%%%%%%%%%%%

\end{document}